\documentclass[pra,amsmath,groupedaddress,floatfix,twocolumn]{revtex4-1}

\usepackage{graphicx}
\usepackage{amssymb}
\usepackage{longtable}
\usepackage{rotating}
\usepackage{array}
\usepackage{multirow}
\usepackage{makecell}
\usepackage{booktabs}
\usepackage{color}

\begin{document}

\title{Efficient systematic scheme to construct second-principles
  lattice-dynamical models}

\author{Carlos Escorihuela-Sayalero,$^{1}$ Jacek C. Wojde\l,$^{2}$ and
  Jorge \'I\~niguez$^{1,2}$}

\affiliation{$^{1}$Materials Research and Technology Department,
  Luxembourg Institute of Science and Technology (LIST), 5 avenue des
  Hauts-Fourneaux, L-4362 Esch/Alzette, Luxembourg\\ $^{2}$Institut de
  Ci\`encia de Materials de Barcelona (ICMAB-CSIC), Campus UAB, 08193
  Bellaterra, Spain}
  
\begin{abstract}
We start from the polynomic interatomic potentials introduced by
Wojde\l\ {\sl et al.} [J. Phys. Condens. Matt. {\bf 25}, 305401
  (2013)] and take advantage of one of their key features -- namely,
the linear dependence of the energy on the potential's adjustable
parameters -- to devise a scheme for the construction of
first-principles-based ({\em second-principles}) models for
large-scale lattice-dynamical simulations. Our method presents the
following convenient features. The parameters of the model are
computed in a very fast and efficient way, as it is possible to recast
the fit to a training set of first-principles data into a simple
matrix diagonalization problem. Our method selects automatically the
interactions that are most relevant to reproduce the training-set
data, by choosing from a pool that includes virtually all possible
coupling terms, and produces a family of models of increasing
complexity and accuracy. We work with practical and convenient
cross-validation criteria linked to the physical properties that will
be relevant in future simulations based on the new model, and which
greatly facilitate the task of identifying a potential that is
simultaneously simple (thus computationally light), very accurate, and
predictive. We also discuss practical ways to guarantee that our
energy models are bounded from below, with a minimal impact on their
accuracy. Finally, we demonstrate our scheme with an application to
ferroelastic perovskite SrTiO$_{3}$, which features many non-trivial
lattice-dynamical features (e.g., a phase transition driven by soft
phonons, competing structural instabilities, highly anharmonic
dynamics) and provides a very demanding test.
\end{abstract}


\maketitle

\section{Introduction}

Over the past decades first-principles simulation methods have
undergone a fantastic transformation, from a rather specialized
technique reserved to a few groups to a standard and powerful tool
accessible to any research team in the world. In particular, methods
based on efficient schemes like Density Functional Theory (DFT)
\cite{hohenberg64,kohn65} make it possible to run predictive
simulations of many materials and systems \cite{martin-book2004}.
However, because of the finite computer power, DFT studies are still
limited to relatively small simulation boxes of a few hundreds of
atoms; further, except in the simplest of cases, computing statistical
averages in realistic conditions is all but impossible. These
limitations can be partly overcome by introducing effective models,
with parameters computed from first principles, that permit faster and
larger-scale calculations (e.g., statistical or dynamical, including
thousands of atoms) while retaining, to some extent, first-principles
predictive power and accuracy \cite{sinnott12}. In the following we
refer to such approaches as {\sl second-principles} methods, adopting
the terminology coined in Ref.~\onlinecite{garciafernandez16}.

While progress in second-principles simulations has been impressive,
this is still an emerging field, and the present situation is somewhat
comparable to that of DFT methods about three decades ago. The
effective models and corresponding simulation tools are reasonably
well established in some particular contexts, as e.g. for the study of
proteins and other biological systems \cite{kukol-book2015}. However,
we are still far from having general tools that permit a
non-specialist to tackle a new problem. The current challenge is to
bring the second-principles techniques to the level that DFT
simulation has reached.

This is a daunting endeavor, much complicated by the fact that, by
their own nature, different second-principles methods may rely on
utterly diverse approximations. Even if we restrict ourselves to
atomistic schemes, we can distinguish between approaches that treat
explicitly both atoms and electrons, and those in which only the atoms
are retained in the description. Among the latter, we can find a
wealth of options to handle the interatomic interactions, ranging from
physically or chemically motivated models (which range from the very
simple -- as the Lennard-Jones \cite{jones24} or bond-valence
\cite{brown09,perezmato04,shin05} potentials -- to the quite complex
-- as e.g. the so-called {\sl ReaxFF} force fields \cite{vanduin01})
to abstract mathematical approaches (e.g., those based on neural
networks \cite{jovanjose12}). Different schemes present different
strengths and weaknesses, and adapt better to different problems. For
example, a suitable method to tackle phenomena that involve formation
and breaking of chemical bonds (e.g., ReaxFF) may not be the most
appropriate to handle relatively simpler situations (e.g., those in
which the bonding topology is preserved) that may nevertheless require
a high quantitative accuracy.

Here we focus on the latter category, i.e., on cases in which an
underlying lattice of chemical bonds can be assumed to remain
essentially unperturbed throughout the simulation. While this {\em
  constant-topology} condition may look restrictive at first, there
are many all-important properties that are perfectly compatible with
it, including most lattice-dynamical phenomena, lattice thermal
transport, dielectric and piezoelectric responses, and even structural
phase transitions driven by soft phonon modes. Wojde\l\ {\sl et
  al}. \cite{wojdel13} took advantage of this constant-topology
assumption to introduce simple and general polynomic potentials that
describe the dependence of the energy -- written as a Taylor series
around a suitable reference structure -- on arbitrary (though
relatively small) atomic distortions. Indeed, in
Ref.~\onlinecite{wojdel13} some of us showed that such potentials
provide a very flexible framework to fit accurately a training set of
first-principles data. In particular, this approach allowed us to
resolve the temperature-driven phase transitions of prototypical
ferroelectric (PbTiO$_{3}$ or PTO) and ferroelastic (SrTiO$_{3}$ or
STO) perovskite oxides \cite{wojdel13}; further, it has allowed us to
discover effects -- ranging from novel structural phases under elastic
constraints \cite{wojdel13}, to the occurrence of ferroelectric phase
transitions within the ferroelectric domain walls of PTO
\cite{wojdel14a}, or the negative-capacitance behavior of the PTO
layers of PTO/STO superlattices \cite{zubko16} -- that had been
essentially missed by previous effective models or direct DFT
simulations. Hence, the methods of Ref.~\onlinecite{wojdel13} have
amply proven their usefulness and deserve further development.

In this article we report our recent work based on the approach of
Ref.~\onlinecite{wojdel13}, and introduce key developments in the way
in which the models are constructed. The original work \cite{wojdel13}
was based on a somewhat primitive (perfectly standard) algorithm that
required us to choose the interaction terms of the model before
fitting the parameters to a training set of DFT data; then, the
fitting process involved successive (and computationally costly)
constrained optimizations based on the minimization of non-linear goal
functions. This scheme is good enough to produce useful models, but it
is far from optimal. In particular, it does not take advantage of some
unique features of our potentials that permit a far more efficient and
powerful approach.

Here we introduce a model-construction procedure that automatically
identifies the most relevant coupling terms and computes the
corresponding parameters. The most important terms are selected from a
pool that may be defined to include virtually all possible interatomic
interactions, of the order of hundreds in the applications considered
in this work. Our scheme is designed to produce {\sl a family} of
models of increasing complexity and accuracy; thus, it makes it
possible to choose the potential better suited to specific
investigations. (Naturally, the computational burden of the ensuing
statistical or dynamical simulations grows with the complexity or size
of the potential.) Such an exhaustive model-construction procedure
relies on a very fast and efficient algorithm for the calculation of
the model parameters: In essence, we recast the usual (costly and
cumbersome) optimization problem into one that involves solving
(almost instantaneously) a system of linear equations. This critical
step is possible thanks to a distinct feature of our polynomial
models, namely, that the energy depends linearly on the parameters to
be fitted. Thus, in summary, our new method makes it possible to
obtain models of exceedingly high accuracy in a fast and robust way.

The paper is organized as follows. In Section~II we introduce the
formalism behind the new model-construction scheme, and also address
issues concerning cross-validation and energy boundedness. In
Section~III we describe the construction of models for STO, a
challenging material to test our method. Different possibilities to
construct a training set and fit the model parameters, as well as
other practical aspects, are discussed in some detail. Finally, in
Section~IV we summarize our work.

\section{Formalism}
\label{sec:formalism}

In this section we describe our approach in a material-independent
way. First we review the model-potential scheme of
Ref.~\onlinecite{wojdel13}, recalling the basic formulas and
notation. Then we describe our present strategy for an automatic model
construction, which takes advantage of the linearity of the potential
with respect to the fitting parameters. Finally, we introduce the
problem of cross-validating the models, and also mention briefly the
question of how to obtain potential energies that are bounded from
below; these practical issues will be better explained with an example
in Section~\ref{sec:application}.

\subsection{Second-principles lattice-dynamical models}

The models proposed by Wojde\l\ {\sl et al.} \cite{wojdel13} can be
described as a Taylor series of the energy, around a certain reference
structure (RS), as a function of atomic displacements (${\bf u}$) and
strains (${\boldsymbol \eta}$). It is convenient to split the energy
$E({\bf u}, {\boldsymbol \eta})$ in three parts:
\begin{equation}
E({\bf u}, {\boldsymbol \eta}) = E_{\rm RS} + E_{\rm p}({\bf u}) +
E_{\rm s}({\boldsymbol \eta}) + E_{\rm sp}({\bf u}, {\boldsymbol
  \eta}) \, ,
\label{Energy_split}
\end{equation}
where the subscripts ``p'', ``s'' and ``sp'' stand for ``phonon'',
``strain'', and ``strain-phonon'', respectively, and $E_{\rm RS}$ is
the energy of the reference configuration. We assume the typical case
in which the RS is a critical point of the PES, so that the
first-order terms of the Taylor series vanish. Further, following
Ref.~\onlinecite{wojdel13}, we write the energy in terms of
displacement differences, so that our potential is explicitly
compliant with the acoustic sum rule at all orders of the
expansion. Using Latin letters to label the atoms in our material and
Greek letters to label Cartesian coordinates, we can write the
following general expression for $E_{\rm p}$:
\begin{widetext}
\begin{equation}
E_{\rm p}({\bf u}) = \frac{1}{2} \sum_{\underset{\alpha\beta}{ijkh}}
\widetilde{K}^{(2)}_{ij\alpha kh\beta} (u_{i\alpha} - u_{j\alpha})
(u_{k\beta} - u_{h\beta})+ \frac{1}{6}
\sum_{\underset{\alpha\beta\gamma} {ijkhrt}}
\widetilde{K}^{(3)}_{ij\alpha kh\beta rt\gamma} (u_{i\alpha} -
u_{j\alpha}) (u_{k\beta} - u_{h\beta}) (u_{r\gamma} - u_{t\gamma}) +
... \; ,
\label{EP}
\end{equation}
\end{widetext}
where the $\widetilde{\bf K}^{(n)}$ tensors can be related with the
$n$-th-derivatives of the energy,
\begin{equation}
K^{(n)}_{ijk...\alpha\beta\gamma...} = \frac{\partial^{n} E}{\partial
  u_{i\alpha}\partial u_{j\beta} \partial u_{k\gamma} ...} \; ,
\end{equation}
by expanding the displacement-difference products in
Eq.~(\ref{EP}). Analogously, we write the strain-phonon term as
\begin{widetext}
\begin{equation}
E_{\rm sp}({\bf u}, {\boldsymbol \eta}) = \frac{1}{2} \sum_{a}
  \sum_{ij\alpha} \widetilde{\Lambda}^{(1,1)}_{aij\alpha} \eta_a
  (u_{i\alpha} - u_{j\alpha}) + \frac{1}{6} \sum_a
  \sum_{\underset{\alpha\beta}{ijkh}}
  \widetilde{\Lambda}^{(1,2)}_{aij\alpha kh\beta} \eta_a
  (u_{i\alpha}-u_{j\alpha}) (u_{k\beta} - u_{h\beta}) + ... \; ,
\label{ESP}
\end{equation}
\end{widetext}
where $\widetilde{\boldsymbol\Lambda}^{(m,n)}$ is the coupling tensor
of order $m$ in strain and $n$ in the atomic displacements, and we use
Latin letters ($a$, $b$, etc.) to label strain components in Voigt
notation \cite{nye-book1985}. Finally we have
\begin{equation}
E_{\rm s}({\boldsymbol \eta}) =
  \frac{N}{2} \sum_{ab} C_{ab}^{(2)} \eta_a \eta_b +
  \frac{N}{6} \sum_{abc}C^{(3)}_{abc} \eta_a \eta_b \eta_c + ... \; ,
\label{ES}
\end{equation}
where ${\bf C}^{(m)}$ is the bare elastic tensor of order $m$. Note
that when working with insulators (as in the example that will be
described below), it is convenient to further split the interactions
involving phonons in long-range (dipole-dipole) and short-range parts
\cite{wojdel13,gonze97}.

The expression for the energy gives us access to all relevant
lattice-dynamical quantities; in particular, the forces on the atoms
are given by
\begin{equation}
f_{i\alpha} = -\frac{\partial E}{\partial
  u_{i\alpha}}\Bigg\rvert_{{\bf u},{\boldsymbol \eta}}
\end{equation}
for a certain $({\bf u},{\boldsymbol \eta})$ configuration, while the
stresses acting on the cell are
\begin{equation}
 \sigma_{a} = -\frac{\partial E}{\partial\eta_{a}}\Bigg\rvert_{{\bf
     u},{\boldsymbol \eta}} \; .
\end{equation}
Note that, in order to match the usual definition of stress, this
derivative needs to be computed under the condition that the {\em
  relative} positions of the atoms in the cell are kept constant. This
is not direct in our scheme, as we work with absolute atomic
displacements ${\bf u}$; nevertheless, in practice this derivative
calculation can be easily tackled by implementing an appropriate chain
rule.

\subsection{Calculation of model parameters}

There is a standard procedure to compute the parameters of a
lattice-dynamical model, directly applicable to any potential type,
including ours. The objective is to obtain a model that reproduces a
{\em training set} (TS) of relevant lattice-dynamical and structural
data. To quantify the model's accuracy in reproducing the TS, one
introduces a positively defined {\em goal function} (GF), and turns
the parameter fitting into a GF minimization problem. Solving this
problem is typically a hard task, as the GF is usually a
high-dimension non-linear function of the free parameters, with an
intricate multi-minima landscape associated to it. Hence, a
time-consuming numerical solution is mandatory in most cases. In
essence, this is the approach some of us adopted in
Ref.~\onlinecite{wojdel13}.

Here we show that, because of the particular form of our interaction
potentials, we can make a judicious choice of GF that permits an
analytic solution to the parameter-fitting problem. This is a drastic
simplification that allows for very fast parameter calculations, which
in turn makes it possible to implement an automatic procedure to
identify the best model from essentially all possible couplings.

\subsubsection{Definitions and goal function}

As it is obvious from the formulas above, our models are linear in the
parameters to be fitted. Hence, we can formally write the energy of a
model with $p$ parameters as
\begin{equation}
E[\Theta_{p}]({\bf u}, {\boldsymbol \eta}) = \sum_{\lambda}
\theta_{\lambda} t_{\lambda}({\bf u}, {\boldsymbol \eta})
+ E^{\rm fixed}({\bf u},  {\boldsymbol \eta}) \; ,
\label{eq:compactSPLD}
\end{equation}
where $\Theta_{p} := \{\theta_{1}, ..., \theta_{p}\}$ is the set of
free parameters and $\mathcal{T}_{p} := \{t_{1}, ..., t_{p}\}$ gathers
the corresponding polynomial terms. Note that for a given parameter
$\theta_{\lambda}$, the corresponding $t_{\lambda}$ includes {\em all}
the symmetry-related terms (i.e., all products of atomic displacements
and strains) whose coupling is given by $\theta_{\lambda}$. We call
$t_{\lambda}$ a symmetry-adapted term (SAT), using the terminology of
Ref.~\onlinecite{wojdel13}. For convenience, we also introduce an
energy $E^{\rm fixed}$ that gathers all the terms that do not need to
be fitted. $E^{\rm fixed}$ will typically include interactions that
can be computed from first principles in a straightforward way, such
as the harmonic bare elastic constants, the harmonic dipole-dipole
couplings, etc.

Now we introduce a goal function $G[\Theta_{p}]$ that (1) is aimed at
constructing models that give a good description of lattice-dynamical
properties and (2) is a simple function of the model parameters. It is
most natural and convenient to choose a GF of the form
\begin{equation}
\begin{split}
 G[\Theta_{p}, {\rm TS}] = \frac{1}{M_{1}} \sum_{s\tau} \left( f^{\rm
   TS}_{\tau}(s)- f_{\tau}[\Theta_{p}](s) \right)^{2} + \\
\frac{1}{M_{2}} \sum_{s a} \Omega^{2}(s) \left( \sigma^{\rm TS}_{a} (s)
- \sigma_{a}[\Theta_{p}](s) \right)^{2}
\end{split} ,
\end{equation}
where $s$ labels the $M$ configurations in the TS and we have
introduced the bijective mapping $i\alpha\leftrightarrow\tau$ to
alleviate the notation. Note that we mark the target forces and
stresses with a ``TS'' superscript, and we also indicate the
parametric nature of the quantities derived from the model. $M_{1}$
and $M_{2}$ are normalization factors computed as the cardinal of the
elements of the corresponding sums. $\Omega(s)$ is the factor that
Sheppard {\sl et al.} \cite{sheppard12} proposed, in the context of
nudged-elastic-band calculations, to properly weight forces and
stresses; it is defined as
\begin{equation}
\Omega(s) = \left(V(s) \sqrt{N}\right)^{-1/3} \; ,
\end{equation}
where $N$ is the number of atoms in the simulation cell and $V(s)$
the cell volume for configuration $s$.

\subsubsection{Analytic minimum of the goal function}

Let us denote the SAT derivatives with respect to atomic displacements
and strains by
\begin{equation}
\bar{f}_{\lambda\tau}(s) = - \frac{\partial t_{\lambda} ({\bf u},
   {\boldsymbol \eta})}{\partial u_{\tau}} \Bigg\rvert_{s}
\end{equation}
and
\begin{equation}
\bar{\sigma}_{\lambda a} = - \frac{\partial t_{\lambda} ({\bf u},
  {\boldsymbol \eta})}{\partial \eta_{a}}\Bigg\rvert_{s} \; ,
\end{equation}
respectively. Then, the forces and stresses computed as derivatives of
Eq.~(\ref{eq:compactSPLD}) can be written as
\begin{equation}
 f_{\tau}(s) =
 \sum_{\lambda} \theta_{\lambda} \bar{f}_{\lambda\tau}(s) +
 f^{\rm fixed}_{\tau}(s)
\end{equation}
and
\begin{equation}
\sigma_{a}(s) = \sum_{\lambda} \theta_{\lambda}
\bar{\sigma}_{\lambda a}(s) + \sigma^{\rm fixed}_{a}(s) \; ,
\end{equation}
respectively, where $f^{\rm fixed}(s)$ and $\sigma^{\rm fixed}(s)$ are
the corresponding derivatives of $E^{\rm fixed}$ evaluated at
configuration $s$. Hence, the GF can be rewritten as
\begin{widetext}
\begin{equation}
 G[\Theta_{p}, {\rm TS}] = \frac{1}{M_1} \sum_{s\tau} \left( f^{\rm
   TS}_{\tau}(s) - \sum_{\lambda} \theta_{\lambda}
 \bar{f}_{\lambda\tau}(s) - f^{\rm fixed}_{\tau}(s) \right)^{2}
+ \frac{1}{M_2} \sum_{sa} \Omega^{2}(s) \left( \sigma^{\rm
  TS}_{a}(s) - \sum_{\lambda} \theta_{\lambda}
\bar{\sigma}_{\lambda a}(s) - \sigma^{\rm fixed}_{a}(s)\right)^{2}  \; .
\end{equation}
\end{widetext}
The extrema of the goal function satisfy $\partial G[\Theta_{p}, {\rm
    TS}]/\partial \theta_{\mu} = 0$ $\forall\,\mu$. This translates
into the set of conditions
\begin{widetext}
\begin{equation}
\frac{2}{M_1} \sum_{s\tau} \left( f^{\rm TS}_{\tau}(s) -
\sum_{\lambda} \theta_{\lambda} \bar{f}_{\lambda\tau}(s) -
f^{\rm fixed}_{\tau}(s) \right) \bar{f}_{\mu\tau}(s)
+\frac{2}{M_2} \sum_{sa} \Omega^{2}(s) \left(\sigma^{\rm TS}_{a}(s) -
\sum_{\lambda} \theta_{\lambda} \bar{\sigma}_{\lambda a}(s) -
\sigma^{\rm fixed}_{a}(s) \right)\bar{\sigma}_{\mu a}(s) = 0 \; ,
\label{nabla1}
\end{equation}
\end{widetext}
for $\mu$ = 1, ..., $p$. This expression can be recast in the
following, more convenient form
\begin{widetext}
\begin{equation}
\begin{split}
\sum_{\lambda} \left[ \sum_{s} \left( \frac{1}{M_1} \sum_{\tau}
  \bar{f}_{\mu\tau}(s) \bar{f}_{\lambda\tau}(s) + \frac{1}{M_2}
  \sum_{a} \Omega^{2}(s) \bar{\sigma}_{\mu a}(s) \bar{\sigma}_{\lambda
    a}(s) \right) \right] \theta_{\lambda} = \\
\sum_{s} \left(
\frac{1}{M_1} \sum_{\tau} \left[ f_{\tau}^{\rm TS}(s)
- f_{\tau}^{\rm fixed}(s) \right] \bar{f}_{\mu\tau}(s) + \frac{1}{M_2}
\sum_{a} \Omega^{2}(s) \left[ \sigma_{a}^{\rm TS}(s)
- \sigma^{\rm fixed}_{a}(s) \right] \bar{\sigma}_{\mu a}(s) \right) \; ,
\label{nabla2}
\end{split}
\end{equation}
\end{widetext}
which we can further simplify to write
\begin{equation}
\sum_{\lambda} \Delta_{\mu\lambda} \theta_{\lambda} = \Gamma_{\mu} \; ,
\label{SOEcompact}
\end{equation}
where $\Delta_{\mu\lambda}$ and $\Gamma_{\mu}$ are trivially defined
by comparing Eqs.~(\ref{nabla2}) and (\ref{SOEcompact}). Hence, we can
find the extrema of $G[\Theta_{p}]$ by solving a system of $p$ linear
equations.

Our GF can be viewed as a $p$-dimensional parabola, and we always have
$G[\Theta_{p}, {\rm TS}] \geq 0$. Further, as shown in
Appendix~\ref{sec:app1}, the eigenvalues of the associated Hessian
($H_{\lambda\mu} = \partial^{2}G/\partial \theta_{\lambda} \partial
\theta_{\mu}$) are either positive or zero. It is thus clear that a
critical point of $G$ must be a minimum, never a saddle point or
maximum. (Note that, by definition, a saddle point requires the
presence of both positive and negative eigenvalues.)

It is also relatively easy to show (see Appendix~\ref{sec:app1}) that,
for a specific choice of $\Theta_{p}$ and TS, we can have two possible
scenarios: either Eq.~(\ref{SOEcompact}) has a single solution or it
has infinite ones. The former case corresponds to the situation in
which there exits a well-defined collection of values of the $\Theta_{p}$
parameters yielding an optimum compromise to reproduce the TS
data. (All the Hessian eigenvalues are positive in this case, and the
solution of Eq.~(\ref{SOEcompact}) is a minimum of $G$.) The latter
case corresponds to the situation in which the solution of
Eq.~(\ref{SOEcompact}) is a manifold with dimension greater than zero,
implying that some linear combinations of parameters can take
arbitrary values without affecting $G$. Such combinations correspond
to Hessian eigenvectors with zero eigenvalue; $G$ does not have a
single minimum, but a manifold of minima, in this case.

\begin{figure}
\includegraphics[width=\columnwidth]{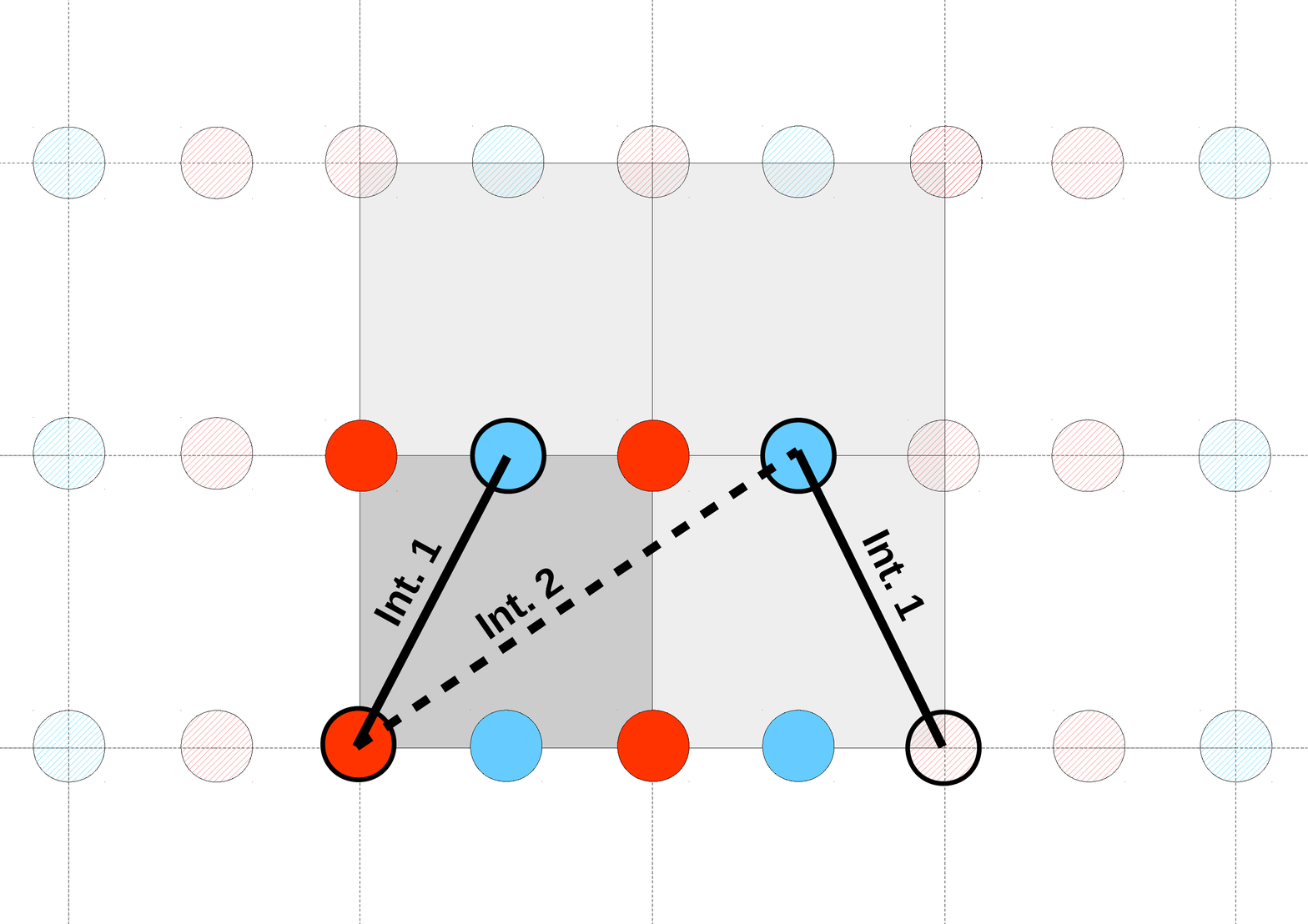}
\caption{Sketch illustrating how the choice of a supercell for the
  generation of the TS data implicitly imposes a spatial cutoff for
  the interatomic interactions. The elemental unit cell of the crystal
  is colored in dark grey, while the supercell used for the TS
  calculations is in light grey. The elemental cell contains two
  atoms. The atoms in the supercell are displayed in strong colors,
  while their periodic images are shaded. The interactions labeled as
  ``Int. 1'' and ``Int. 2'' are obviously different, but they cannot
  be resolved on the basis of data obtained for this supercell. Note
  that the longest-range ``Int. 2'' connects atoms that are also
  linked by ``Int. 1'' due to the periodic supercell repetition.}
\label{fig:equivalent-interactions}
\end{figure}

It is interesting to note the reasons why there may be linear
dependencies in our system of equations. Let us consider
$\Delta_{\mu\lambda}$ and $\Gamma_{\mu}$ in Eq.~(\ref{SOEcompact});
these are $p$-dimensional objects that encapsulate all the information
in the TS in a compact way. In the limit of a large $p$ and a
relatively small TS, it may happen that the TS does not provide enough
information to fit all the parameters in $\Theta_{p}$. In such a case,
the system of equations will be underdetermined, yielding infinite
solutions. A similar difficulty appears when the TSs are obtained in
ways that only explore a subset of the configuration space, e.g., most
typically, by running molecular dynamics (MD) simulations in
relatively small supercells. In such a case, the periodic boundary
conditions associated with the employed supercell effectively define
the spatial range of the interatomic couplings that can be
resolved. Indeed, as far as the description of the TS data is
concerned, and as sketched in Fig.~\ref{fig:equivalent-interactions},
interactions extending beyond that supercell-defined distance become
exactly equivalent to other shorter-range couplings. Hence, they lead
to linear dependencies when setting up the system of equations in
Eq.~(\ref{SOEcompact}).

In practice, it is easy to handle the situations in which
Eq.~(\ref{SOEcompact}) has infinite solutions. On one hand, if we
believe our TS is complete enough, the presence of infinite solutions
suggests that we can simplify our model without loss of accuracy. We
can do so in an orderly and physically-motivated way, e.g., by
removing from $\Theta_{p}$ the longest-range interactions that lead to
linear dependencies in Eq.~(\ref{SOEcompact}). On the other hand, if
we are interested in quantifying precisely all the interactions in our
model, the presence of infinite solutions indicates that we need to
extend the TS so as to remove the linear dependencies. In this case,
we will typically need to include information that is qualitatively
different from that in the original TS, e.g., obtained from MD
simulations of larger supercells, etc.

\subsubsection{Finding the best model of $p$ terms}

We are interested in constructing models that are accurate and, at the
same time, permit fast calculations. Hence, we need to find a way to
construct the simplest (computationally lightest) models that
reproduce the TS with a certain accuracy. Naturally, we can use the
goal function as the measure of accuracy, so that the above problem
translates into finding the simplest model whose associated GF is
below a certain threshold. (As we will see in
Section~\ref{sec:application}, in order to evaluate a model's accuracy
and predictive power, we will eventually adopt a practical approach
that goes beyond evaluating the GF.)

Let ${\mathcal T}_{P}$ be the set of $P$ terms, with associated
parameters $\Theta_{P}$, that define {\em all} the possible
interactions in our material of interest. Given the specific form of
our interatomic potentials, ${\mathcal T}_{P}$ can be easily defined
by three approximations or cutoffs: the maximum order of the
polynomial, the maximum spatial range of the interactions, and the
maximum number of bodies in the interaction terms. For specific
choices of these three cutoffs, one can implement an algorithm that
identifies all the SATs the model can potentially contain. We will
typically have a very large number of them, of the order of $P$~=~500
in the application discussed below.

Let us use the expression {\em $p$-model} to refer to a model of $p$
terms, ${\mathcal T}_{p} \subset {\mathcal T}_{P}$ and $\Theta_{p}
\subset \Theta_{P}$ being the corresponding sets of terms and
parameters, respectively. We need a way to find the best $p$-model,
i.e., the choice of ${\mathcal T}_{p}$ and $\Theta_{p}$ that minimizes
the GF when we restrict ourselves to models of $p$ terms. A
brute-force approach to this search -- by computing all possible
$p$-models and comparing the corresponding GF values -- would be a
daunting task. For example, for a representative case of $P = 500$ and
$p = 20$, we have about 10$^{35}$ different models. Obviously, in
spite of our efficient strategy to compute the parameters for an
specific choice of ${\mathcal T}_{p}$, considering so many
possibilities is computationally unfeasible.

We overcome this difficulty by constraining the model search,
implementing what can be described as a {\em stepwise procedure with
  forward selection} \cite{xu01}. In short, we start with $p = 1$ and
identify the best 1-model, a problem that we solve exactly by
considering all $P$ possible candidates. Let ${\mathcal T}_{1}^{*}$
denote the best 1-model. Then, we move to $p+1$ and consider all
possible ${\mathcal T}_{p+1}$ models subject to the constraint that
${\mathcal T}_{p}^{*} \subset {\mathcal T}_{p+1}$. In other words, we
only consider $p+1$-models that contain the terms of the best
$p$-model. We can solve this problem exactly, by considering all $P-p$
possibilities explicitly. Then we iterate the procedure until a
sufficiently small GF is obtained.

We have checked the reliability of the above procedure in two
ways. First, for a number of ${\mathcal T}_{P}$ choices with small
$P$, we run a brute-force search for the best $p$-models, and compare
the exact results thus obtained with the outcomes of our proposed
strategy. In essence, our constrained approach succeeds in identifying
the best $p$-model in almost all cases, and the very few exceptions
correspond to cases with very small values of $p$ (in our work with
STO, we find this problem for $p < 5$, a limit where the corresponding
models are not physically sound anyway). Second, we consider the
following refinement of our algorithm: once ${\mathcal T}_{p}^{*}$ has
been identified, we check whether it is possible to improve the GF by
replacing one of the chosen $p$ terms by one of the remaining
$P-p$. We find that such a refinement seldom improves the best
$p$-model, and the few cases in which an improvement is observed
correspond to very small $p$ values.

\subsection{Cross-validation}
\label{sec:cross}

In Statistics, {\em cross-validation} is a common procedure to analyze
the predictiveness of a model, and is often used as {\em stopping
  criterion} in model construction methodologies.

A rather usual approach to it is the so-called leave-$n$-out method
\cite{celisse08}. This method consist in the following steps. Given a
TS, we remove from it $n$ randomly selected elements, and fit the best
$p$-model to the remaining elements by minimizing the GF. Let $G(p)$
denote the resulting GF value. Then, we test the accuracy of such best
$p$-model by evaluating the GF using the $n$ TS elements left out of
the fit. Let $G^{\rm test}(p)$ denote the resulting GF
value. Naturally, we expect $G(p) > G(p+1)$ and $G^{\rm test}(p) >
G^{\rm test}(p+1)$, as models should improve as they become more
complex. Then, if we find $p^{\rm best}$ such that $G^{\rm
  test}(p^{\rm best}) < G^{\rm test}(p^{\rm best}+1)$, this indicates
that our model is loosing predictive power as it grows, and we can
thus identify the optimum size ($p = p^{\rm best}$) of the model.

Unfortunately, in actual applications this procedure tends to render
$G^{\rm test}(p)$ curves that display several minima, making it
difficult to identify the most predictive model \cite{prechelt98}. In
our particular case, as we detail below, we find a different
complication: We work with well-populated TSs constructed from MD
trajectories, and the TS elements (i.e., the atomic configurations
representative of the trajectory) tend to contain similar information;
hence, our calculated $G^{\rm text}(p)$ curves do not present any
minimum. As we will see below, we resolve this difficulty by creating
a physically-motivated test set that allows us to perform a convincing
cross-validation and verify that our models will be reliable for the
calculation of properties of interest.

\subsection{Energy boundedness}
\label{sec:boundedness1}

Before concluding this section, we should note the main weakness of
the approach just described, namely, that the energy of our optimum
models is likely to be unbounded from below.

Our scheme involves running an unconstrained search for the best
$p$-model, testing polynomial terms from a pool $\mathcal{T}_{P}$ that
essentially includes all possible couplings compliant with cutoffs
concerning the order of the expansion, the spatial range of the
interactions, and the number of bodies in the interacting terms. The
behavior of the model for large atomic distortions is dominated by the
highest-order terms, which are not guaranteed to be positive
definite. In other words, nothing in our construction scheme prevents
the models from presenting run-away solutions with $E \to -\infty$.

Naively, one could expect that a complete enough TS will automatically
produce bounded models. More precisely, high-energy atomic
configurations, involving large distortions of the RS, provide
information about the restoring forces that keep the atoms together;
hence, in order to tackle the unboundedness problem, it may seem
sufficient to include representative configurations of that sort
(e.g., as obtained from high-temperature MD runs) in the
TS. Unfortunately, our experience indicates that this procedure is not
enough to constrain all possible run-away directions in the general
case and, thus, it does not lead to an automatic model-generation
scheme.

In this work we have considered a number of possible,
potentially-automatic solutions to this problem, briefly described
below. These experiments have led us to identify a practical strategy
that, while requiring some inspection of the best models (i.e., while
it is not fully automatic as implemented at present), appears as a
satisfactory compromise, allowing to impose boundedness without any
significant loss of accuracy.

\section{Application to S\lowercase{r}T\lowercase{i}O$_{3}$}
\label{sec:application}

We have chosen SrTiO$_{3}$ as our test material. STO is one of the
most interesting perovskite oxides, and it has been receiving
continued attention for years because of its critical importance to
the field (as, e.g., it is the most widely used substrate on which
thin films of other perovskites are grown \cite{schlom07}) and the
unique physical effects it displays either in combination with other
materials (e.g., exotic two-dimensional electron gas at the interface
of STO with LaAlO$_{3}$ \cite{ohtomo04}, novel ferroelectric effects
at superlattices of STO with PTO \cite{zubko16,bousquet08}) or by
itself (e.g., polar order at the ferroelastic domain walls of STO at
low temperatures \cite{scott12,salje13}). This wealth of interesting
properties is partly due to STO's unique and challenging
lattice-dynamical behavior, which turns this compound into an unique
test case for our automatic potential-construction method. Our
conjecture is that, if our scheme allows us to tackle STO
successfully, it will probably allow us to investigate other,
relatively simpler compounds as well.

\begin{figure}
\includegraphics[width=\columnwidth]{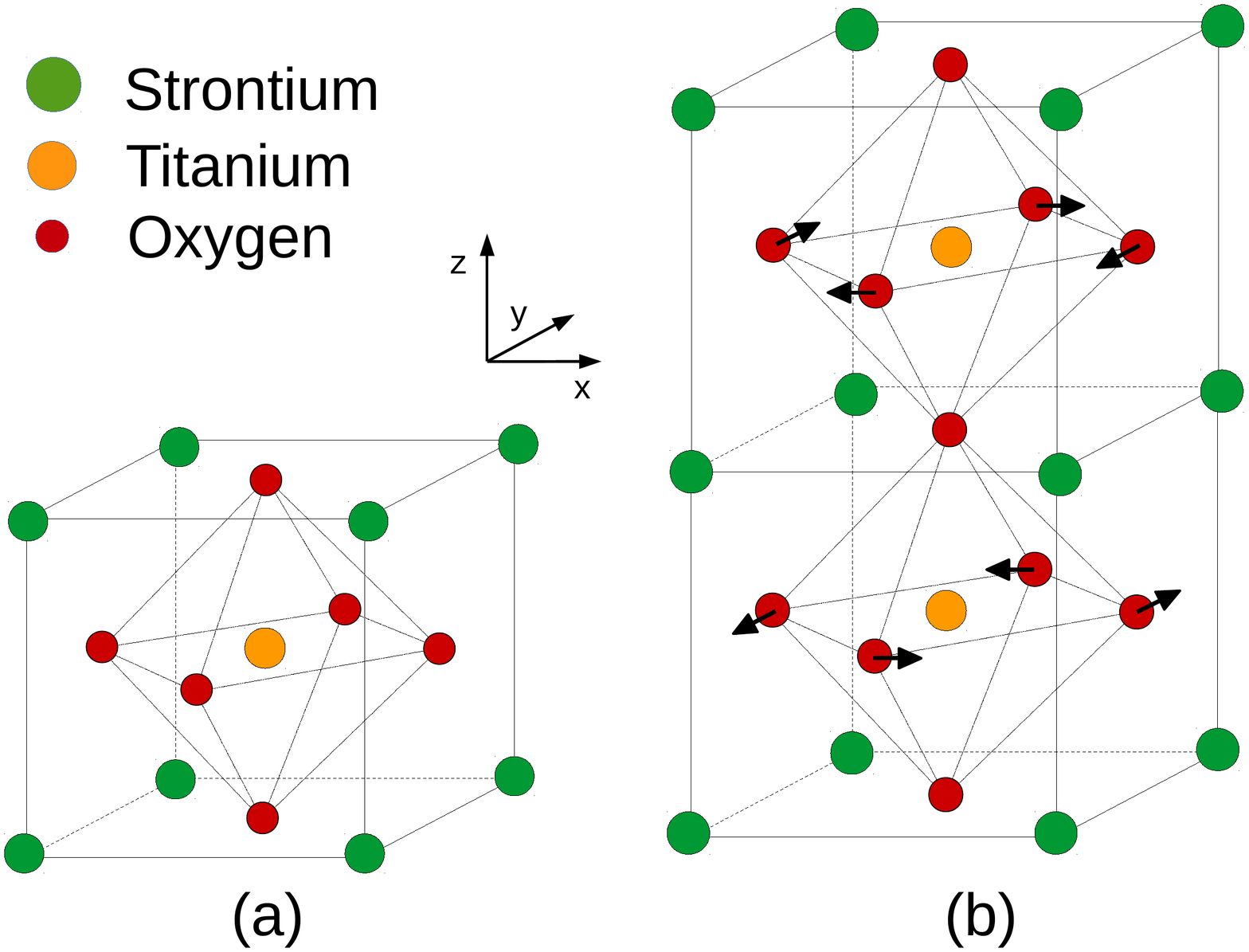}
\caption{Sketch of the STO perovskite structure. Panel~(a) shows the
  elemental cell for the high-symmetry cubic ($Pm\bar{3}m$)
  structure. Panel~(b) sketches the antiphase O$_{6}$ rotations that
  characterize the AFD distortion driving the structural phase
  transition in STO.}
\label{fig:perovs}
\end{figure}

STO crystallizes in the perovskite structure, sketched in
Fig.~\ref{fig:perovs}. It displays a cubic phase (with $Pm\bar{3}m$
space group) at high temperatures [Fig.~\ref{fig:perovs}(a)], but
transforms into a so-called antiferrodistortive (AFD) structure
[Fig.~\ref{fig:perovs}(b)] when the temperature falls below $T_{\rm
  C}$~=~110~K \cite{lytle64,fleury68}. This low-temperature phase has
a tetragonal symmetry ($I4/mcm$), featuring concerted rotations of the
oxygen octahedra about one of the principal axes of the perovskite
lattice. Such rotations are modulated in antiphase when we move from
cell to cell along the direction of the rotation axis, and the
resulting structure is usually labeled $a^{0}a^{0}c^{-}$ in the
notation introduced by Glazer \cite{glazer72}. [In the following we
  will assume that the rotation axis lies along the $z$ direction, as
  shown in Fig.~\ref{fig:perovs}(b).]

Another interesting feature of STO is its {\em quantum paraelectric}
character \cite{muller79}. On top of the mentioned AFD structural
instabilities (soft phonon modes) that drive the transition between
the cubic and tetragonal phases, experiments \cite{muller79} and
first-principles theory \cite{zhong96} indicate that the cubic phase
of STO presents yet another unstable distortion, namely, a polar mode
whose condensation would create a macroscopic electric
polarization. The reason why this spontaneous polarization is not
observed experimentally in bulk STO samples is two-fold. On one hand,
such a ferroelectric (FE) soft mode competes with the AFD instability,
and the occurrence of the latter tends to stabilize the former
\cite{zhong95b}. On the other hand, quantum fluctuations prevent the
condensation of the weakened FE instability \cite{zhong96}. The
presence of such quantum effects (which rely on the wave-like
character of the relatively-heavy atoms in STO) has been amply
demonstrated theoretically and experimentally, their most obvious
fingerprint being the anomalous behavior of the dielectric constant of
STO at low temperatures (see Ref.~\onlinecite{muller79} for
experiments on STO, and Ref.~\onlinecite{akbarzadeh04} for simulation
results for another representative quantum paraelectric, KTaO$_{3}$).

The present work focuses on the construction of effective potentials
that accurately reproduce the PES of a material as obtained from first
principles. Hence, a detailed comparison of the predictions of such
potentials with the experimental observations is secondary in this
context. In the case of our application to STO, we will solve our best
models by means of {\em classical} Metropolis Monte Carlo (MC)
simulations \cite{binder-book2010}, with the purpose of verifying that
we obtain the correct qualitative behavior and investigating how
details of the potential affect an all-important feature, i.e., the
phase-transition temperature $T_{\rm C}$. However, while we will
briefly comment on the relation between our results and the known
facts about STO, a quantitative comparison with experiment is not
pertinent. Indeed, because our simulations are classical, they do not
include the quantum effects that strongly affect the behavior of STO
at temperatures around and below the phase transition. A more detailed
investigation of STO's properties, as predicted by our models and in
connection with experiment, remains for future work.

\subsection{Details of the first-principles calculations}

To generate our TS data, we use DFT within the local density
approximation (LDA) as implemented in the {\sc VASP} package
\cite{kresse96}. The choice of LDA over other energy functionals is
not trivial. On one hand, it is known that LDA overbinds, predicting
equilibrium volumes that are generally smaller than the experimental
ones by 1-2\%, and that this error can have a dramatic influence in
the PES of materials like STO, which are very reactive to applied
stresses. Hence, the LDA may not be the most appropriate choice if we
are aiming at reproducing STO's experimental behavior with
quantitative accuracy. On the other hand, there is an ample literature
on STO modeling, including the construction of coarse-grained
effective potentials \cite{zhong96,zhong95a}, based on LDA
results. Hence, using the LDA allows us to make a more direct
comparison with those previous theoretical works -- including the STO
model that some of us developed in Ref.~\onlinecite{wojdel13} --,
which is the most important consideration in the context of this work.

Further details of our LDA calculations are as follows. We treat the
ionic cores using the projector-augmented wave method
\cite{blochl94,kresse99}, and solve explicitly for the following
electrons: Sr's 3$s$, 3$p$, and 4$s$; Ti's 3$s$, 3$p$, 4$s$, and 3$d$;
and O's 2$s$ and 2$p$. The electronic wave functions are represented
in a plane-wave basis truncated at 500~eV. Brillouin zone integrals
are computed in a grid of 6$\times$6$\times$6 $k$-points for
calculations involving the 5-atom perovskite cell as
periodically-repeated unit, or equivalent meshes when larger
supercells are used. We have checked that these calculation conditions
render results that are sufficiently accurate for our purposes.

Structural minimizations are run until residual force components are
below 0.01~eV/\AA. Langevin molecular dynamics simulations are run
using the Parrinello-Rahman scheme \cite{allen-book1989,parrinello81},
to allow fluctuations of the cell volume and shape at zero applied
pressure. The Langevin parameters are chosen to obtain a fast approach
to the targeted temperature and reduce fluctuations. (Since we are not
interested in obtaining converged equilibrium properties, the details
of the Langevin dynamics are not critical.) The MD runs are always
initialized with random velocities, and the atomic motions are
unconstrained (thus, no symmetries are preserved). We typically use a
time step of 2~fs.

\subsection{Definition of the best possible model}
\label{sec:pool}

As done in Ref.~\onlinecite{wojdel13}, and customary in theoretical
works on ferroelectric and dielectric perovskites
\cite{zhong95b,zhong94a,zhong95a,bellaiche00,kornev07}, we choose as
our RS the cubic phase of the material, as obtained from a
symmetry-constrained structural relaxation using the LDA. The high
symmetry of this phase (full cubic $Pm\bar{3}m$ group) drastically
reduces the number of allowed independent couplings in our polynomial
expansion, which simplifies the task of parameter fitting.

The relevant PES of STO is characterized by relatively small atomic
distortions; indeed, when we compare the cubic (RS) and tetragonal
(ground state) structures, we find that the maximum change in bond
distances is about 0.16~\AA, corresponding to Sr--O
pairs. Additionally, we know that the cubic phase is unstable against
various structural distortions (most notably, the AFD modes driving
the transition to the low-temperature structure), which implies that
the harmonic part of the PES is not bounded from below. Given these
facts, we decided to consider a 4-th order Taylor series as the
simplest possible theory that is physically sound for STO, as in
particular it permits energy boundedness from below.

As regards the purely elastic part of the model, we adopt the simplest
possible approximation and include only harmonic terms. This choice
seems both accurate (as the strains involved in STO's structural phase
transition are relatively small, below 0.5~\%) and sufficient to
produce a physically sound model. Note that STO's cubic phase does not
present any purely elastic instability \cite{wojdel13}, and the
harmonic elastic energy is thus bounded from below. As regards the
strain-phonon coupling energy, we restrict ourselves to the
lowest-order terms allowed by symmetry, which are quadratic in the
displacements and linear in the strains.

We split the interatomic interactions in two parts, short-range
(chemical) and long-range (Coulombic), the latter being amenable to
analytical treatment. More precisely, we treated the long-range
electrostatic couplings exactly as in Ref.~\cite{wojdel13}: We used
the well-known analytical formula for the harmonic coupling between
electric dipoles, which ultimately depends only on the atomic Born
charge tensors and the high-frequency (purely electronic) dielectric
response of the material \cite{gonze97}. As regards the short-range
harmonic interactions, previous investigations on the phonon band
structure of the cubic phase of STO \cite{wojdel13} and similar
compounds \cite{ghosez99} show that they decay quickly with distance,
and that a 2$\times$2$\times$2 repetition of the elemental 5-atom cell
is enough to capture all the significant ones. As regards the
short-range anharmonic interactions, previous modeling works
\cite{wojdel13} indicate that they are even shorter in range. (They
are treated as on-site or self-energy terms in most models in the
literature \cite{zhong95a,waghmare97,krakauer99,bellaiche00}.)

Hence, we consider all the terms within the interaction range defined
by the 2$\times$2$\times$2 40-atom supercell. We allow anharmonic
interactions to extend as far as the short-range harmonic ones, i.e.,
up to the maximum distance effectively defined by our 40-atom
supercell. Finally, we allow all our coupling terms to involve as many
atoms (bodies) as their order allows.

We then generate all the symmetry-allowed polynomial terms compatible
with these cutoffs. For $E_{\rm p}$ we obtain 45 harmonic terms, 79
3rd-order ones, and 275 4th-order ones; for $E_{\rm s}$ we obtain 3
terms, and 161 terms for $E_{\rm sp}$. Note that these terms only
depend on the structure and symmetry of the ideal cubic perovskite
phase; hence, they can be applied to the study of any such material,
not only STO.

In the following we present various exercises aimed at exploring our
new recipe to fit the lattice potentials thus defined from a TS of DFT
data. Yet, two parts of these potentials (the bare elastic energy and
the dipole-dipole interactions, both truncated at the harmonic level)
can be trivially obtained from perturbational \cite{gonze97,wu05} or
finite-difference DFT simulations. Further, these two parts of the
energy are very simple, and would not benefit from any additional
optimization provided by our systematic fitting procedure. Hence, for
these terms we use the values directly obtained from DFT and given in
Appendix~\ref{sec:app2}; they are thus included in $E^{\rm fixed}$
[Eq.~(\ref{eq:compactSPLD})] in the fits.

\subsection{Preliminary test: exact harmonic potential}
\label{sec:harmonic}

The harmonic part of our model describes the PES around the RS, i.e.,
the interatomic couplings determining the phonon spectrum of the cubic
phase. The phonon spectrum of a given structure can be computed from
first principles via perturbative \cite{gonze97} or
finite-displacement methods that give access to the interactions in
real space. Hence, it should be relatively easy for us to construct
models of the type proposed here and having an essentially exact
harmonic part; note that the construction approach described in
Ref.~\onlinecite{wojdel13} took advantage of this fact.

However, the present scheme focuses on fitting, as opposed to
explicitly computing, the model parameters to produce potentials of
optimal complexity and size. Further, here we work with a Taylor
series written in terms of products of displacement differences, the
harmonic part of our potential having the form
\begin{equation}
E_{\rm p}^{\rm har}({\bf u}) = \frac{1}{2}
\sum_{\underset{\alpha\beta}{ijkh}} \widetilde{K}^{(2)}_{ij\alpha
  kh\beta} (u_{i\alpha} - u_{j\alpha}) (u_{k\beta} - u_{h\beta}) \, ,
\end{equation}
in contrast with the usual expression in terms of products of simple
displacements
\begin{equation}
E_{\rm p}^{\rm har}({\bf u}) = \frac{1}{2}
\sum_{\underset{\alpha\beta}{ijkh}} K^{(2)}_{ij\alpha\beta}
u_{i\alpha} u_{j\beta} \, ,
\end{equation}
which is commonly used in phonon analysis. Hence, it is worth testing
our present scheme by applying it to obtain the harmonic part of the
PES around the RS, to explicitly confirm whether our fitting procedure
and displacement-difference representation are able to yield an
essentially perfect description.

To do this, we use a TS composed of the following structures: the
LDA-relaxed cubic phase of STO in a 2$\times$2$\times$2 supercell (our
RS) and slightly distorted versions of the RS in which we move
individual atoms by 0.015~\AA. (These are exactly the same structures
one would consider to compute the phonons of STO's cubic phase by the
finite-displacements method.) Then, we run the fitting procedure
described in Section \ref{sec:formalism}, considering a set
$\mathcal{T}^{\rm harm}_{P}$ that includes all the possible (45)
short-range harmonic terms compatible with the cutoffs described
above. As mentioned above, the harmonic dipole-dipole interactions are
computed analytically and included in $E^{\rm fixed}$.

To present our results, let us recall that the harmonic part of the
energy can be expressed in a more compact form if we use the basis of
eigenvectors of the force-constant matrix ${\bf K}^{(2)}$. Thus, given
a model of $p$ harmonic terms $\mathcal{T}^{\rm har}_{p}$, we can
expand the displacement-different products to obtain the corresponding
force-constant matrix ${\bf K}^{(2)}[\Theta]$, diagonalize it, and
compare the results with the exact LDA solution.

\begin{figure}
\includegraphics[width=\columnwidth]{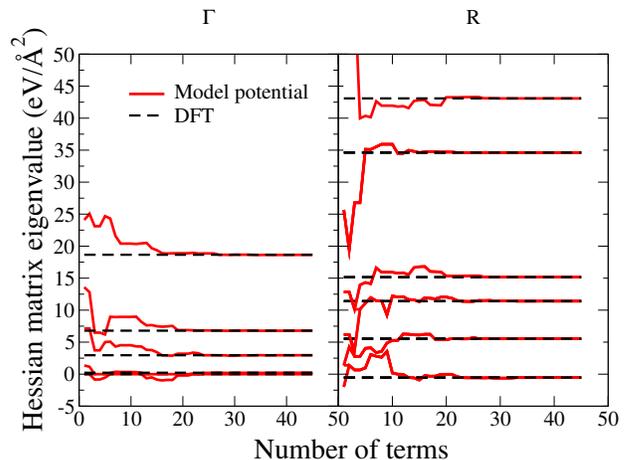}
\caption{Eigenvalues obtained from the diagonalization of the harmonic
  part of the phonon energy $E_{\rm p}$, at two specific $q$-points
  $\Gamma$ and $R$ (see text). The solid lines indicate the results
  obtained from the fitted model as a function of the number of terms
  included, while the exact DFT-computed results are given by dashed
  lines.}
\label{fig:harmonic}
\end{figure}

Figure~\ref{fig:harmonic} shows the results as a function of the
number of terms in the harmonic model; in particular, we show the
eigenvalues corresponding to $q$-points $\Gamma$ [${\bf q}_{\Gamma} =
  (0,0,0)$] and $R$ [${\bf q}_{R} = 2\pi/a (1/2,1/2,1/2)$, where $a$
  is the lattice constant of the 5-atom RS cell]. As we can see, the
agreement is essentially perfect for $p$-models with $p \gtrsim 30$,
confirming that our scheme is able to produce exact harmonic models
{\em and}, in passing, automatically identify the dominant
interactions. In Section~\ref{sec:analysis} we will briefly discuss
the most important of such couplings.

\subsection{Choice of training sets, targeted models}

We generate several TSs aimed at exploring the configuration space
accessible at various temperatures. More precisely, we generate TSs at
10~K and 300~K; additionally, we compute test sets at 200~K and
500~K. Our training and test sets are obtained from MD runs employing
a 40-atom (2$\times$2$\times$2) STO supercell, which, as argued above,
we deem sufficient to obtain information about the relevant
short-range interactions. Further, the MD runs are performed at 0~GPa,
allowing the cell volume and shape to fluctuate during the simulation;
this permits access to information concerning the strain-dependent
parts of the potential.

Note that, strictly speaking, our small 40-atom simulation supercell
does not allow us to talk about a well defined temperature in our MD
runs; our target temperature is just a convenient handle that we use
to push the system into exploring different regions of configuration
space.

\begin{figure}
\includegraphics[width=\columnwidth]{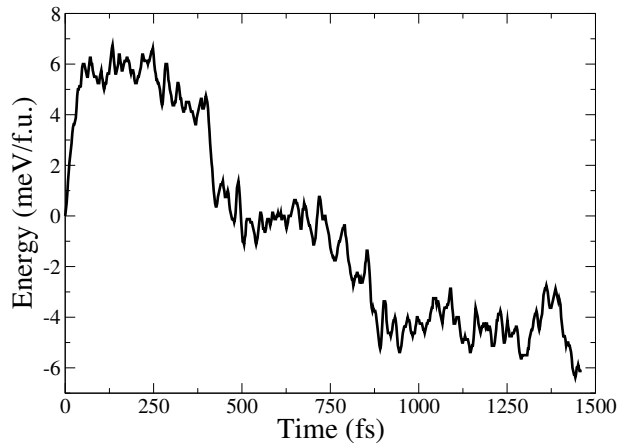}
\caption{Energy evolution as a function of time corresponding to the
  typical MD run used to construct our TSs of DFT data. This
  particular case corresponds to the run used to build TS@10 (see
  text). The run starts from the RS, which is taken to be the zero of
  energy. Eventually the simulated system finds its way to the ground
  state structure and fluctuates around it.}
\label{fig:MD-example}
\end{figure}

We find it important to start all our MD runs from the RS, for the
following reason. As shown in Fig.~\ref{fig:MD-example}, the length of
our typical MD run is about 1500~fs, including an initial stage in
which the system explores the surroundings of the cubic phase (the
potential energy increases considerably as the atoms move following
the random initial velocities compatible with a 10~K kinetic energy)
to eventually (at about 400~fs in the figure) start moving towards the
ground state. Once a MD run is computed, we construct the
corresponding TS by picking 100 configurations from the trajectory; we
pick such configurations randomly, but making sure they are
homogeneously distributed throughout the MD trajectory, so as to
retain information on the initial part of the run. This is necessary
to obtain models that yield accurate energy differences. Indeed,
because our scheme is based on fitting energy derivatives, in order to
get an accurate energy difference between two structures (e.g., and
most importantly, between the RS and the ground state), we need the
model to describe well the connecting path.

Here we discuss models constructed by fitting to three different TSs,
one obtained from a MD run at 10~K (denoted ``TS@10'' in the
following), a second one obtained from a MD run at 300~K (``TS@300''),
and a third one that combines configurations from TS@10 and TS@300
(``TS@10+300''). The TS@10 explores the low-energy configurations of
the material, and contains the kind of information that has been
traditionally included in effective models of STO and similar
compounds \cite{zhong95b,zhong94a,zhong95a,bellaiche00,kornev07}. {\sl
  A priori} we do not expect the models obtained from TS@10 to
accurately describe the configuration space that the material explores
at higher temperatures; hence the interest in considering TS@300 as
well. Finally, we consider TS@10+300 to investigate the possibility of
creating models that give a good description of low- and high-energy
configurations simultaneously.

As we will see below, obtaining an accurate description of the
low-energy PES of STO is not an easy task. In particular, we find it
critical to complement the TSs with a single additional configuration
that competes in energy with the ground state but is not visited in
our default MD runs. Such a low-energy state is denoted
$a^{0}a^{0}c^{+}$ in Glazer's notation; it involves in-phase rotations
of the O$_{6}$ octahedra about one of the principal axes of the
perovskite lattice (recall that the rotations are in antiphase in
STO's ground state). Hence, our TS@10 and TS@10+300 training sets are
completed with this additional piece of information.

Finally, as shown in Section~\ref{sec:harmonic}, it is possible to
compute the harmonic part of the model in an essentially exact
way. Hence, here we discuss models in which we fit all the short-range
phonon and strain-phonon couplings freely (``free models'' or FMs in
the following) and also models in which the exact harmonic part of
$E_{p}$ is retained in $E^{\rm fixed}$ (``exact harmonic models'' or
EHMs in the following). This allows us to test the importance (and
evaluate the convenience) of having a perfect description of the
harmonic energy around the the RS.

Table~\ref{tab:model-list} lists the six models that we discuss below,
indicating the short labels we use to denote them.

\begin{table}
\caption{List of the STO models constructed in this work (see
  text). The number of parameters in the optimum cross-validated
  models is indicated in the last column. For the EHMs, we also
  indicate that we retain 45 harmonic terms in $E_{\rm p}$, i.e., all
  the independent interactions within our $2\times2\times2$
  supercell.}
\vskip 2mm
\begin{tabular}{cccc}
\hline\hline
Number & Fit type & TS & Parameters retained \\
\hline
1 & FM  & TS@10     & 33 \\
2 & EHM & TS@10     & 45+10 \\
3 & FM  & TS@300    & 37 \\
4 & EHM & TS@300    & 45+14 \\
5 & FM  & TS@10+300 & 44 \\
6 & EHM & TS@10+300 & 45+17 \\
\hline\hline
\end{tabular}
\label{tab:model-list}
\end{table}

\subsection{Finding optimum models}

Given a TS and a pool of possible parameters, the fitting procedure
described in Section~\ref{sec:formalism} is automatic. Yet, one
critical issue remains, namely, how to determine the optimal size of a
model. The search for optimum models comprises two aspects. At a
practical level, we want to identify the simplest (smallest) models
that are computationally light and, at the same time, reproduce the TS
data accurately enough. At a more basic level, we want to produce
models that are predictive, i.e., that do not suffer from {\sl
  overfitting}. To achieve this latter goal, we must implement a
cross-validation procedure, which will be the focus of the following
discussion.

\begin{figure}
\includegraphics[width=\columnwidth]{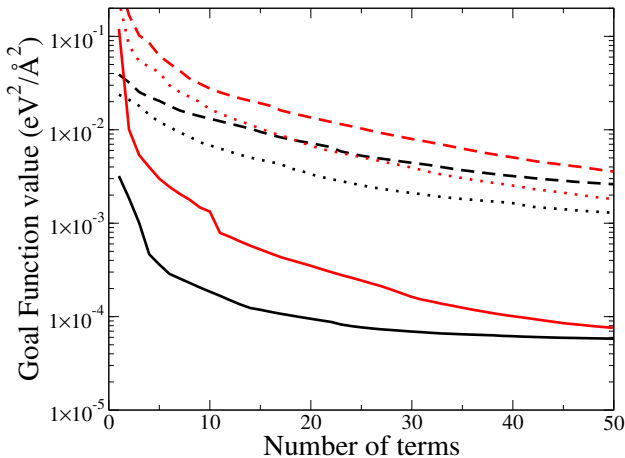}
\caption{Optimized goal function, as a function of the number of
  terms, for the various models considered in this work. Red and black
  lines correspond to FM and EHM models, respectively. Solid, dashed,
  and dotted lines correspond to models fitted to the TS@10, TS@300,
  and TS@10+300, respectively. See text for details.}
\label{fig:GF}
\end{figure}

\subsubsection{Classic cross-validation}

Figure~\ref{fig:GF} shows the evolution of the goal function $G$, as a
function of the number of terms in the model, for each of the six
cases in Table~\ref{tab:model-list}. In all cases the GF evolves
rather smoothly and converges, in the limit of a large number of
parameters, to values in the range between
10$^{-4}$~eV$^{2}$/\AA$^{2}$ and 10$^{-2}$~eV$^{2}$/\AA$^{2}$. At
convergence, the lowest GF values correspond to the two models fitted
to TS@10, while the largest ones correspond to the two TS@300
potentials. This indicates that it is harder to get accurate fits of
MD data obtained at higher temperatures. (Since the higher-temperature
TSs involve larger distortions of the RS, we should be able to improve
the accuracy of the TS@300 and TS@10+300 models by increasing the
maximum order of our polynomic potential.) Finally, Fig.~\ref{fig:GF}
shows that, for a given TS and same number of fitted parameters, the
exact harmonic models are always more accurate than the free
models. This was to be expected, as EHM fits start from a nearly
perfect description of the energetics of small distortions (involving
45 harmonic parameters fitted as in Section~\ref{sec:harmonic}), while
the FM fits start from scratch. As suggested by the results in
Fig.~\ref{fig:GF}, the EHM and FM curves corresponding to a particular
TS tend to merge in the limit of very large models.

As the models increase in size, the fits rely more and more on details
of the TS data that may not relate to the basic lattice-dynamical
behavior of the material; instead, they may be the result of small
intricate features of the interaction potential, or even be caused by
numerical inaccuracy in the DFT simulations. Models incorporating that
kind of information will loose in predictive power, an undesired
effect that is usually known as {\em overfitting}. As a first attempt
to determine the optimum size of a predictive model, we apply the
cross-validation procedure described in Section ~\ref{sec:cross} --
i.e., the leave-$n$-out method --, which we slightly modify for
convenience.

\begin{figure}
\includegraphics[width=\columnwidth]{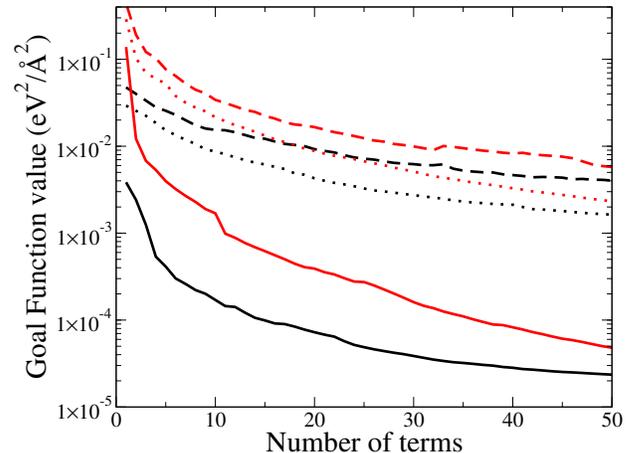}
\caption{Behavior of the goal function evaluated using fitted models
  and test-set configurations, as described in the text. The colors
  and line types are as in Fig.~\ref{fig:GF}.}
\label{fig:CV1}
\end{figure}

To explain how we proceed, let us consider the representative case of
the FMs fitted to TS@10. We take the family of FMs fitted to the
complete TS@10, whose corresponding GF curve is shown in
Fig.~\ref{fig:GF}, as our models to cross-validate. Then, we resort to
the MD run from which we obtain TS@10, and simply select a different
collection of 50 configurations to construct a test set that is
qualitatively similar to TS@10. Then, we evaluate the GF by using the
FM parameters fitted to TS@10 {\em but} summing over the
configurations in the test set. The results for the GF thus evaluated,
as a function of the number of parameters in the model, are shown in
Fig.~\ref{fig:CV1} for all six models in
Table~\ref{tab:model-list}. Notably, the obtained curves are
qualitatively (and quantitatively) similar to those of
Fig.~\ref{fig:GF}, which were obtained from actual GF
minimizations. (For the TS@10 models, the GF curves in
Fig.~\ref{fig:CV1} are lower than those in Fig.~\ref{fig:GF}, which we
have checked is a coincidence related to our selecting -- randomly --
a relatively {\sl benign} test set.) Most importantly, the GF curves
in Fig.~\ref{fig:CV1} decrease monotonically, i.e., they do not
display any minimum and do not allow us to determine an optimum model
size \cite{prechelt98}. Further analysis suggest that this {\em
  failure} is related to the fact that our TSs are quite complete and,
hence, the leave-$n$-out method is not testing the predictive power of
the corresponding models in any significant way. Noting that, in
general, we will be able to afford (and want to work with) exhaustive
TSs as the ones considered here, the leave-$n$-out cross-validation
method seems inadequate for our purposes.

\begin{figure}
\includegraphics[width=\columnwidth]{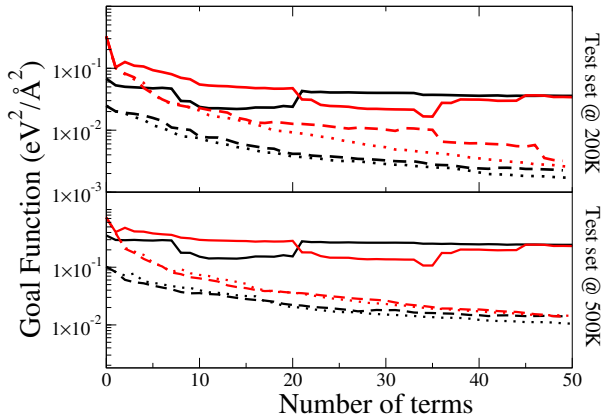}
\caption{Behavior of the goal function evaluated using fitted models
  and test-set configurations, as described in the text. The colors
  and line types are as in Fig.~\ref{fig:GF}.}
\label{fig:CV2}
\end{figure}

\subsubsection{Alternative cross-validation criteria}
\label{sec:cv-alternative}

We then try to validate our models against DFT data that are
qualitatively different from those used to construct them. Our first
experiment is to use test sets obtained from MD trajectories
thermalized to two temperatures (namely, 200~K and 500~K) that deviate
significantly from those of our reference TSs (i.e., 10~K and
300~K). The results for the GFs evaluated against such test sets are
shown in Fig.~\ref{fig:CV2}. Most interestingly, the curves
corresponding to the TS@10 models are not trivial in this case: they
present one or more minima, which could provide us with a criterion to
select optimum predictive potentials. It is worth noting that, when we
test the TS@10 models against 500~K data, such minima correspond to a
relatively small number of parameters; in contrast, the minima occur
for relatively large models when we use the test set obtained at
200~K. This reflects the fact that the TS@10 models become less
accurate for the description of larger RS distortions, as those
corresponding to MD trajectories obtained at higher temperatures;
accordingly, the cross-validation procedure suggest that simpler
TS@10-fitted potentials will do a better job at capturing the general
features of typical high-temperature configurations of the material.

Figure~\ref{fig:CV2} also shows that the GF values for the TS@10
models are relatively large; in contrast, those for the TS@10+300 and
TS@300 models are quite small, and approach those in Fig.~\ref{fig:GF}
obtained from an explicit GF optimization. This shows that the
TS@10+300 and TS@300 models excel at describing the configurations
associated to the 200~K and 500~K MD trajectories. (Actually, even the
performance of the TS@10 model is much better than it may seem from
Fig.~\ref{fig:CV2}, as we will see below.) Accordingly, these test
sets are not challenging the predictive power of such models, and the
corresponding GF curves are monotonically decreasing. Hence, this
cross-validation strategy is not fully satisfactory either.

\begin{figure*}
\includegraphics[width=2.0\columnwidth]{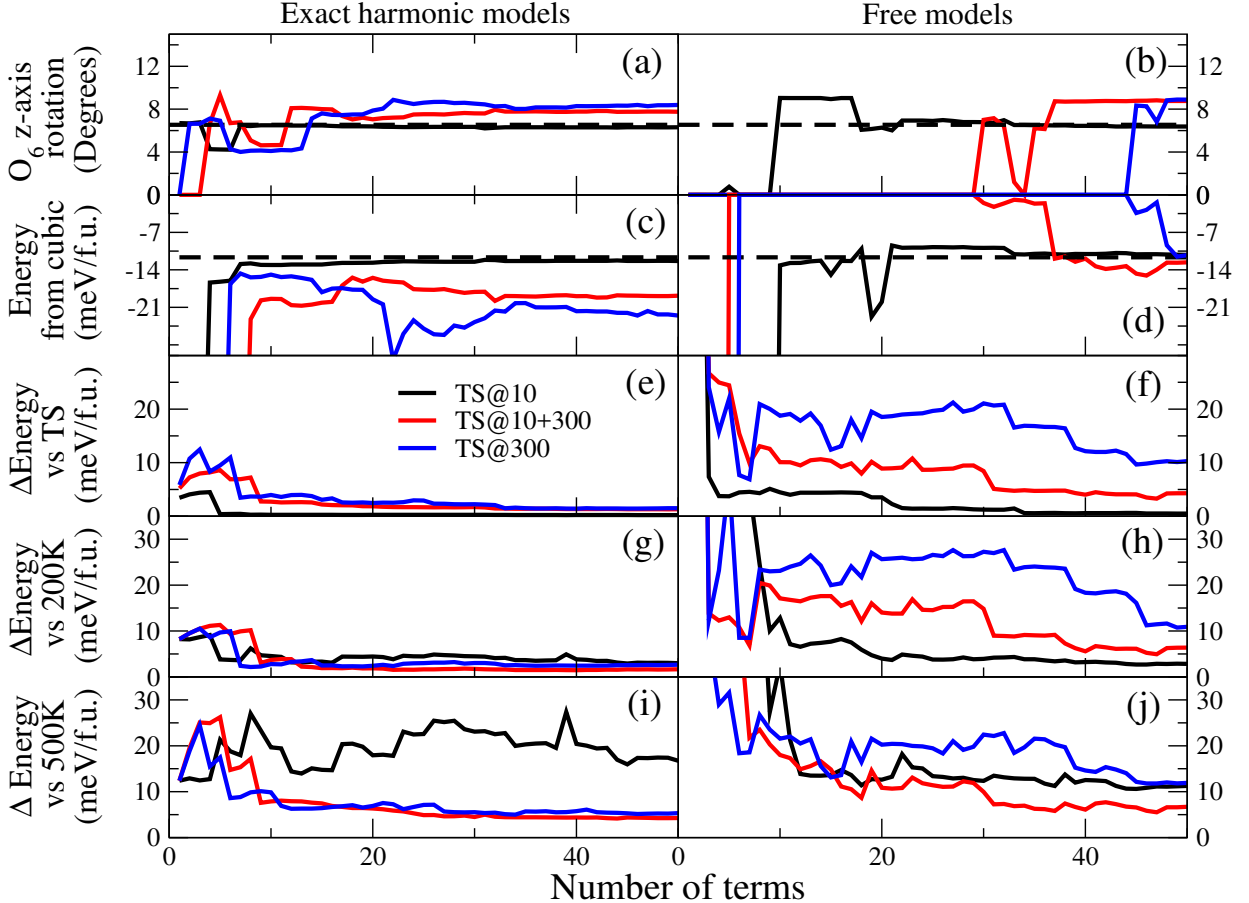}
\vskip -8mm
\caption{Performance of the six models discussed in this work as
  regards the prediction of various ground state [panels (a-d)] and MD
  [panels (e-j)] related quantities. To obtain the data in panels
  (e-j), we use our model potentials to compute the energy of the
  configurations visited in the MD trajectories computed from DFT at
  various temperatures; we report the average difference between the
  DFT energy and that obtained from the model.}
\label{fig:CV3}
\end{figure*}

Finally, we try to cross-validate our models by checking their
predictions for several key structural and energetic features of the
PES of STO. In particular, for materials like STO, which undergoes a
structural phase transition driven by a soft phonon mode, it has been
shown \cite{abrahams68,wojdel14b} that the transition temperature
correlates strongly with the energy difference between the high- and
low-symmetry phases (which are, in our case, the cubic RS and the
tetragonal ground state, respectively) and the amplitude of the
corresponding structural distortion (for STO, this is best quantified
by the rotation angle of the O$_{6}$ octahedra). Figure~\ref{fig:CV3}
thus shows what our models predict for these critical quantities, as a
function of the number of parameters. The figure also shows how well
each particular model reproduces the energetics of the MD trajectory
used to construct the TS to which the model was fitted (thus, e.g.,
the TS@10 models are tested against the MD trajectory obtained at
10~K). Finally, Fig.~\ref{fig:CV3} shows how well our models reproduce
the energetics of the DFT-computed MD trajectories at 200~K and
500~K. Note that our interest in using energy differences to test our
models is two-fold: On one hand, the energy is the critical quantity
that will determine the equilibrium properties of our materials of
interest as a function of temperature, as obtained e.g. from Monte
Carlo simulations. On the other hand, we do not (explicitly) use
energies to calculate the model parameters; hence, checking the
energies computed with our potentials implies a test of their
predictive power, even if we restrict ourselves to the configurations
in the TSs used for the fits.

The results shown in Fig.~\ref{fig:CV3} can be viewed from two
different angles. On one hand, they are somewhat disappointing as a
cross-validation exercise. For example, it could be argued from these
results that the performance of the TS@10 and TS@300 models tends to
deteriorate, for an increasing number of terms, when it comes to
describe the 500~K [e.g. panel~(i)] and ground state [e.g. panel~(c)]
data, respectively. However, it is all but impossible to identify an
optimum model size from the erratic-looking curves that we
obtain. Further, in the TS@10+300 cases, the performance improves
rather continuously, for all the considered quantities, as the models
grow; in other words, our new set of tests is not challenging
sufficiently the predictive power of the TS@10+300 models.

On the other hand, the results in Fig.~\ref{fig:CV3} provide a very
convenient quantitative test of the quality of the models; a test that
is far more useful than the GF results in Figs.~\ref{fig:GF},
\ref{fig:CV1} and \ref{fig:CV2}, whose significance is difficult to
judge. For example, Fig.~\ref{fig:CV3} reveals that the TS@10+300 and
TS@300 models have great difficulties to capture the energy difference
between RS and ground state with acceptable precision, despite the
great quantitative accuracy suggested by the previous GF-based
criteria. (Indeed, note that the FMs for TS@300 and TS@10+300 predict
the correct ground state only after a rather large number of terms
have been included.) In contrast, the data in Fig.~\ref{fig:CV3} shows
that the TS@10 models perform incredibly well to predict the energy of
the configurations that are typical of the 200~K MD trajectory
(average deviations being below a tiny 5~meV/f.u.), and give
acceptable results (with deviations between 10 and 20~meV/f.u.) for
the 500~K structures. Hence, while the TS@10 models may have seemed
quite crude according to the GF results of Fig.~\ref{fig:CV2}, the
data in Fig.~\ref{fig:CV3} indicates that they may constitute the most
reasonable choice among all the options we have investigated.

In conclusion, Fig.~\ref{fig:CV3} provides us with the criteria that
we find most convincing to determine the optimum size of our
models. The results in the figure allow us to quantify the accuracy of
the potentials according to a set of demanding tests that are directly
linked to the properties that will be most relevant for future
statistical and dynamical simulations. Hence, we think they provide us
a sufficient information to determine the minimal models that are both
predictive and sufficiently accurate.

\begin{figure}
\includegraphics[width=\columnwidth]{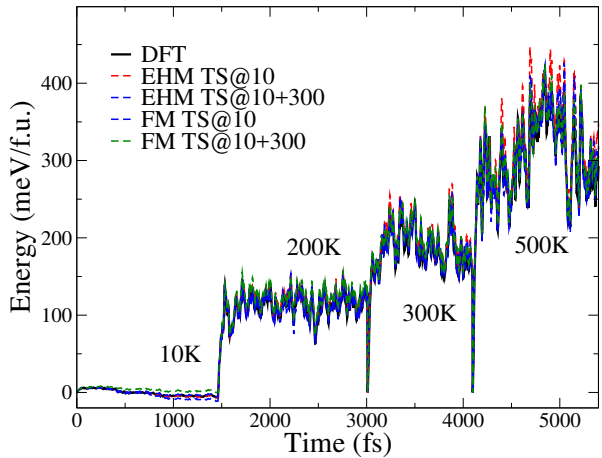}
\caption{Comparison of the DFT-computed energy for MD trajectories at
  various temperatures and the energies obtained by evaluating our
  optimum model potentials for the corresponding structures. Except in
  the 10~K and 500~K cases, the deviations in the energy are all but
  invisible in the scale of the figure, emphasizing the good overall
  accuracy of our fitted models.}
\label{fig:MD-trajectories}
\end{figure}

\subsubsection{Optimum models}

By inspection of Fig.~\ref{fig:CV3}, we decided the optimum size of
the six models of interest here. The number of parameters finally
chosen for each of the models is indicated in
Table~\ref{tab:model-list}. Figure~\ref{fig:MD-trajectories} shows how
well the models reproduce the energy of the configurations in our
DFT-computed MD trajectories, evidencing the great overall
agreement. Note that the energy deviations for a trajectory at a given
temperature are minute when compared with the energy difference
between trajectories at different temperatures; note also that it is
the latter energy scale that is expected to govern the most notorious
dynamic and thermodynamic properties of the material, in particular
its structural transformations. Table~\ref{tab:detail} gives
additional quantitative details on the performance of the TS@10 and
TS@10+300 models. Note that we have not included the results for the
models fitted to TS@300 here, as their poor results for STO's ground
state properties discourages further consideration of such potentials.

\begin{table*}
\caption{Properties characterizing the behavior of the optimum models
  that we consider for MC simulations. Values obtained with the
  original optimum models are given; we also give in parenthesis the
  values computed using the energy-bounded models. In the last row,
  $\Delta E$ is the average energy difference, for the configurations
  included in the corresponding TS, between the LDA results and the
  energies evaluated with our potentials.}
\vskip 2mm
\begin{tabular}{cccccc}
\hline\hline
& FM TS@10 & EHM TS@10 & FM TS@10+300 & EHM TS@10+300 & LDA \\
\hline
O$_{6}$ rotation (degress) &
6.5 (6.5) & 6.5 (6.5) & 9.9 (8.7) & 8.3 (8.1) & 6.5 \\
Ground state energy (meV/f.u.) &
$-$11.1 ($-$11.1) & $-$13.2 ($-$13.1) & $-$13.2 ($-$11.9) & $-$22.9
($-$20.7) & $-$11.7 \\
Goal function (eV$^{2}$\AA$^{-2}\times 10^{-3}$) &
0.14 (0.14) & 0.18 (0.19) & 2.79 (2.81) & 4.94 (5.40) & \\
$\Delta E$ (meV/f.u.) &
0.58 (1.45) & 0.30 (0.27) & 4.74 (4.78) & 2.31 (2.57) & \\
\hline\hline
\end{tabular}
\label{tab:detail}
\end{table*}

\subsection{Analysis of the optimum models, interactions}
\label{sec:analysis}

We now comment on the models produced by our fitting procedure, which
automatically selects the most relevant interactions out of a pool
that virtually includes all possible ones. (Note that we retain an
average of 38 terms in our optimum FM models, out of over 550 possible
ones.) Here we focus on the general aspects that pertain to our
potential-construction approach, touching only briefly on the
particular physics of STO.

\begin{figure}
\includegraphics[width=\columnwidth]{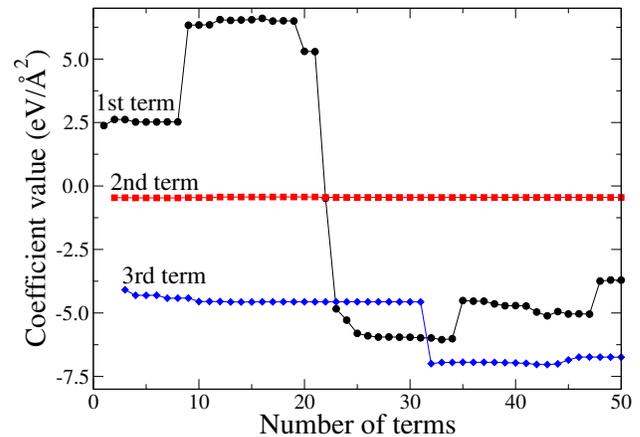}
\caption{Values of the three parameters that our fitting procedure
  selects as most relevant to construct the EHM fitted to TS@10. The
  figure shows their evolution as a function of the total number of
  terms included in the model. Note that all three terms are linear in
  strain and quadratic in the atomic distortions; hence, they all have
  the same units, as indicated in the figure.}
\label{fig:parameters-evolution}
\end{figure}

Our model fitting procedure is essentially a black box, and one may
wonder about the actual physical significance of the specific
interactions and parameter values associated to a specific best
$p$-model. Note that, as we consider increasingly complex models
${\mathcal T}_{p+1}$ with $p+1$ terms, all the $p$ parameters in the
simpler model ${\mathcal T}_{p}^{*} \subset {\mathcal T}_{p+1}$ are
refitted, and their values will inevitably change. Hence, the actual
value of a specific interaction depends on the number of terms in the
model, which raises the question of how significant such a value
actually is. The situation is illustrated in
Fig.~\ref{fig:parameters-evolution}, which shows the evolution of the
three most important terms that are automatically identified when
constructing the EHM fitted to TS@10; the specific couplings
associated to such parameters are indicated in
Table~\ref{tab:parameters-evolution}. As we can see, the second most
important parameter remains nearly constant as we increase the number
of interactions in the model. In contrast, the first and third most
important parameters display drastic changes in their fitted value; in
fact, the interaction determined to be most critical, which describes
the way in which cell strains control the interactions between
nearest-neighboring oxygens (see
Table~\ref{tab:parameters-evolution}), even changes sign as the model
goes above 22 terms, and does not seem to converge to a steady value
even for rather complete models.

\begin{table*}
\caption{Each row corresponds to one of the three parameters
  identified to be the most relevant ones for the EHM model fitted to
  TS@10, indicated by \#1, \#2, and \#3, respectively. We indicate the
  corresponding interaction (in the notation of
  Appendix~\ref{sec:app2}) and, in parenthesis, the value of the GF
  corresponding to the best 1-model, 2-model, and 3-model,
  respectively. Further, we indicate the terms that are related with
  these most important ones and whose inclusion in the model causes
  the discontinuities in their values shown in
  Fig.~\ref{fig:parameters-evolution}. Thus, for parameter \#1, we
  also include parameters \#9 and \#22. In such cases, we give in
  parenthesis the value of the GF that corresponds to considering a
  best 1-model composed of parameter \#9 or \#22; note that these
  values are very close to the minimum GF obtained for parameter
  \#1. All GF values are given in eV$^{2}$\AA$^{-2}$.}
\vskip 2mm
\begin{tabular}{lll}
\hline\hline
\#1: $\eta_2(\text{O1}_{x}-\text{O2}_{x})(\text{O1}_{y}-\text{O2}_{y})$
(0.00320) & 
\#9: $\eta_1(\text{Ti}_{x}-\text{O3}_{x})^2$ (0.00328)&
\#22: $\eta_3(\text{O2}_{y}-\text{O3}_{y})^2$ (0.00324)\\
\#2: $\eta_1(\text{O2}_{z}-\text{O3}_{z})^2$ (0.00185) & & \\
\#3: $\eta_4(\text{Sr}_z - \text{O1}_{z})(\text{Sr}_y-\text{O1}_{y})$
(0.00101) &
\#32: $\eta_4(\text{O1}_{y} - \text{O3}_{y})(\text{O1}_{x} - \text{O3}_{x})$
(0.00102)\\
\hline\hline
\end{tabular}
\label{tab:parameters-evolution}
\end{table*}

Interestingly, the most significant changes in the values of the first
and third parameters in Fig.~\ref{fig:parameters-evolution} occur
rather abruptly, and coincide with the inclusion of very specific
interaction terms in the best $p$-model. The details are summarized in
Table~\ref{tab:parameters-evolution}. We can see that, for example,
the discontinuities in the value of the first parameter (number 1 in
Table~\ref{tab:parameters-evolution}) correspond to the inclusion of
additional strain-phonon couplings (numbers 9 and 22 in the same
table) that bear obvious similarities with the interaction chosen to
be most relevant; naturally, these couplings are connected to the same
kind of interatomic forces and cell stresses in our TS data, and their
computed values are strongly dependent on whether or not we include
all of them in the fit. In other words, the parameter values computed
for the simpler models effectively account for the additional
interactions that are not included, which can be viewed as a
renormalization of sorts. Table~\ref{fig:parameters-evolution} also
gives the values of the GF that determined the selection of a given
interaction term instead of the related ones; thus, for example, when
finding the best 3-model, the parameter picked as number 3 got a GF
score of 0.00101~eV$^{2}$/\AA$^{2}$, while the parameter eventually
picked as number 32 got 0.00102~eV$^{2}$/\AA$^{2}$. Hence, we find
that related interaction terms tend to render similarly good fits, and
retaining one of them instead of the competing ones may well be a
matter of the details of our TS.

These observations clearly suggest that we should be cautious and
avoid overinterpreting the physical relevance of the particular
interaction terms retained in our optimum models, especially in the
case of relatively simple potentials. Obviously, there is physical
information in the couplings automatically identified to be most
relevant (more on this below); however, the specifics depend strongly
on the size of the model and, presumably, the details of the TS.

\begin{figure}
\includegraphics[width=\columnwidth]{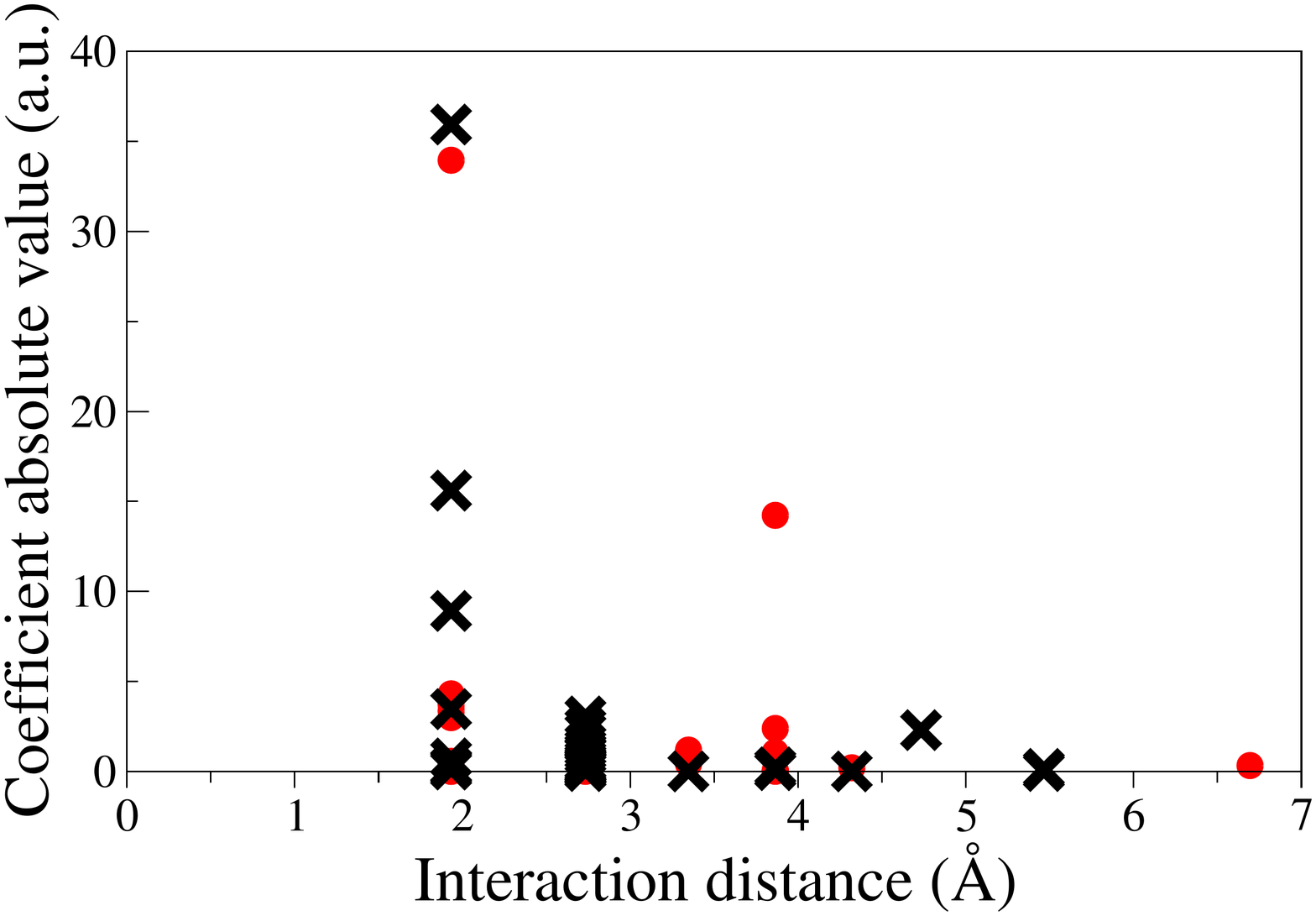}
\caption{Values of the parameters obtained for our FM fitted to TS@10
  and ordered as a function of the longest interatomic separation
  associated to the corresponding coupling. The black crosses
  correspond to the parameters finally retained in the optimum model
  (i.e., to those selected as most relevant by our automatic fitting
  procedure), while we show the coefficients corresponding to the
  disregarded terms with red circles. Representative pairs of
  interacting atoms are indicated in the horizontal axis, following
  the notation in Appendix~\ref{sec:app2}. Note that this figure
  includes information about coefficients corresponding to different
  orders of our Taylor series, which thus have different units; hence,
  we indicate arbitrary units (a.u.) and stress that this figure is to
  be taken only as a qualitative illustration of the spatial decay of
  the interactions.}
\label{fig:parameters-decay}
\end{figure}

Let us now turn to aspects that are more specifically related to our
subject case, STO, focusing on two essential and non-technical
issues. First, when defining our pool of potential interactions in
Section~\ref{sec:pool}, we assumed that our short-range harmonic
couplings, and all anharmonic ones, decay quickly with the interatomic
distance. Figure~\ref{fig:parameters-decay} shows representative
results confirming that such a fast decrease in the magnitude of the
interactions is observed in our fitted models, clearly supporting our
initial hypothesis. Indeed, we find that our optimum models -- which
we determine based on validation criteria that are oblivious of the
nature of the underlying couplings -- turn out to be very
short-ranged; as can be seen in Fig.~\ref{fig:parameters-decay}, all
the retained interactions are within a 5.5~\AA\ range.

\begin{figure}
\includegraphics[width=\columnwidth]{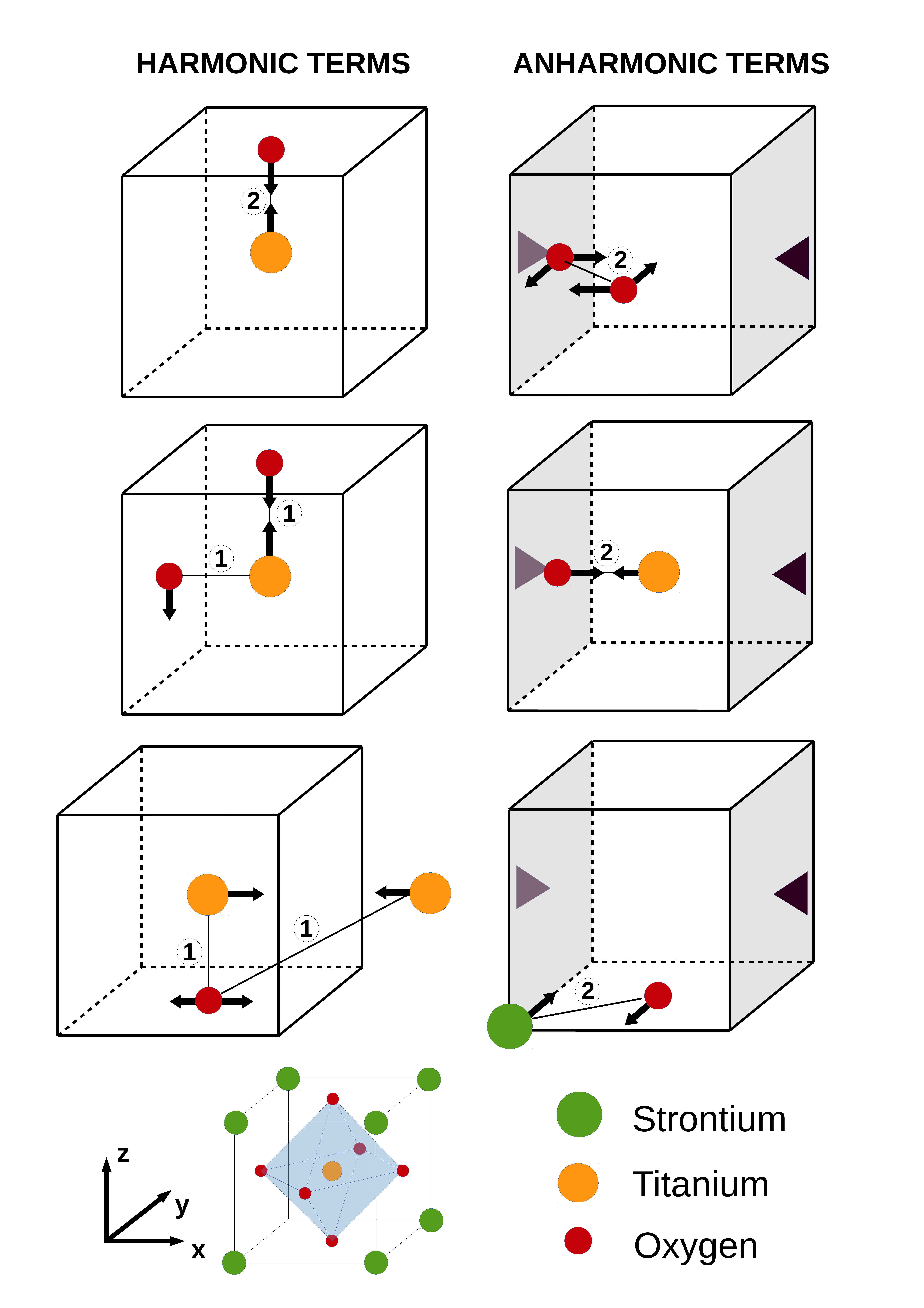}
\caption{Sketch of some of the most important interactions identified
  by our automatic fitting procedure. The atomic displacements
  corresponding to a representative term of the SAT are shown; strains
  are indicated with dark triangles. The circled numbers indicate the
  order of specific displacement-difference terms. Note that all the
  sketched interactions play a role to control the energetics of AFD
  modes and polar distortions, as described in the text.}
\label{fig:main-interactions}
\end{figure}

\begin{table*}
\caption{List of most important interactions, as selected by our
  automatic fitting procedure, for our FM and EHM models fitted to
  TS@10 and TS@10+300. The harmonic part of $E_{\rm p}$ is common to
  both EHM models. The numbers indicate the order in which the
  interactions appeared in the corresponding automatic selection
  process. We include the leading interactions for the different parts
  of the energy in our potentials. The interactions are specified
  following the notation of Appendix~\ref{sec:app2}. We indicate with
  an asterisk the couplings depicted pictorially in
  Fig.~\ref{fig:main-interactions}.}
\begin{tabular}{ll}
\hline\hline\\[-2ex]
\multicolumn{2}{l}{\textbf{Free models}} \\
\multicolumn{1}{l}{\bf TS@10} &
\multicolumn{1}{l}{\bf TS@10+300}\\
\multicolumn{2}{l}{Harmonic part of $E_{\rm p}$}\\
  1*: $(\text{Ti}_y-\text{O2}_{y})^2$
& 1*: $(\text{Ti}_y-\text{O2}_{y})^2$ \\
  2*:
  $(\text{\text{O3}}_{y}-\text{\text{Ti}}_y[010])(\text{\text{Ti}}_y-\text{\text{O3}}_{y})$
& 2*: $(\text{O3}_{y}-\text{Ti}_y[010])(\text{Ti}_y-\text{O3}_{y})$\\
  6: $(\text{Sr}_x-\text{O1}_{x})(\text{Sr}_z-\text{O3}_{z})$
& 4: $(\text{Sr}_x-\text{O1}_{x})(\text{Sr}_z-\text{O3}_{z})$\\
\multicolumn{2}{l}{Anharmonic part of $E_{\rm p}$}\\
  13: $(\text{Sr}_x-\text{O2}_{x})^2(\text{Sr}_z-\text{O3}_{z})$
&  8: $(\text{Ti}_y-\text{O2}_{y})^3$ \\
  14:
  $(\text{Ti}_z[010]-\text{O2}_{z}[010])(\text{Ti}_z-\text{O2}_{z}[010])(\text{Ti}_y-\text{O2}_{y}[010])$
& 14:
  $(\text{Ti}_z[010]-\text{O2}_{z}[010])(\text{Ti}_z-\text{O2}_{z}[010])(\text{Ti}_y-\text{O2}_{y}[010])$\\
  18:
  $(\text{Sr}_x-\text{O2}_{x})^2(\text{O2}_{x}-\text{Sr}_x[001])^2$
& 16: $(\text{Sr}_z-\text{O1}_{z})(\text{Sr}_y-\text{O1}_{y})^2$\\
\multicolumn{2}{l}{Strain-phonon coupling $E_{\rm sp}$}\\
  3*:
  $\eta_2(\text{\text{O}}_{1x}-\text{\text{O2}}_{x})(\text{\text{O1}}_{y}-\text{\text{O2}}_{y})$
& 3*: $\eta_3(\text{Ti}_z-\text{O3}_{z})^2$ \\
  4*: $\eta_2(\text{Sr}_x-\text{O3}_{x})^2 $
& 9*: $\eta_2
  (\text{O1}_{x}-\text{O2}_{x})(\text{O1}_{y}-\text{O2}_{y})$\\
  5:
  $\eta_6(\text{Sr}_y-\text{O3}_{y}[001])(\text{Sr}_x-\text{O3}_{x}[001])$
& 10*: $\eta_2(\text{Sr}_x-\text{O3}_{x})^2$\\[3ex]
\multicolumn{2}{l}{\bf Exact harmonic models}\\
\multicolumn{2}{l}{Harmonic part of $E_{\rm p}$}\\
\multicolumn{2}{l}{1*: $(\text{Ti}_y-\text{O2}_{y})^2$}\\
\multicolumn{2}{l}{2*:
$(\text{Ti}_x-\text{O2}_{x})(\text{Ti}_x-\text{O1}_{x})$}\\
\multicolumn{2}{l}{3*: 
$(\text{O3}_{y}-\text{Ti}_{y}[010])(\text{Ti}_y-\text{O3}_{y})$}\\ [2ex]
\multicolumn{1}{l}{\bf TS@10} &
\multicolumn{1}{l}{\bf TS@10+300} \\
\multicolumn{2}{l}{Anharmonic part of $E_{\rm p}$}\\
  4: $(\text{Ti}_x-\text{O2}_{x})^2(\text{O1}_{y}-\text{O3}_{y})^2$ 
& 2: $(\text{Ti}_y-\text{O2}_{y})^3$ \\
  5:
  $(\text{Ti}_z[010]-\text{O2}_{z}[010])(\text{Ti}_z-\text{O2}_{z}[010])(\text{Ti}_y-\text{O2}_{y}[010])$ 
& 5:
$(\text{Ti}_z-\text{O1}_{z})(\text{Ti}_z-\text{O3}_{z}[001])^2(\text{Ti}_z-\text{O3}_{z})$ 
  \\
  6: $(\text{Sr}_x-\text{O2}_{x})^2(\text{Sr}_z-\text{O3}_{z})$
& 7:
  $(\text{Ti}_z[010]-\text{O2}_{z}[010])(\text{Ti}_z-\text{O2}_{z}[010])(\text{Ti}_y-\text{O2}_{y}[010])$
  \\
\multicolumn{2}{l}{Strain-phonon coupling $E_{\rm sp}$}\\
  1*:
  $\eta_2(\text{\text{O1}}_{x}-\text{\text{O2}}_{x})(\text{\text{O1}}_{y}-\text{\text{O2}}_{y})$ 
& 1*:  $\eta_3(\text{\text{Ti}}_z-\text{\text{O3}}_{z})^2$\\
  2: $\eta_1(\text{\text{O2}}_{z}-\text{\text{O3}}_{z}[100])^2$ 
& 3*:
  $\eta_2(\text{\text{Sr}}_{x}-\text{\text{O3}}_{x})^2$
  \\
  3*:
  $\eta_4(\text{\text{Sr}}_z-\text{\text{O1}}_{z})(\text{\text{Sr}}_y-\text{\text{O1}}_{y})$
& 4: $\eta_1(\text{\text{O3}}_{y}-\text{\text{O1}}_{y}[010])^2$\\[1ex]
\hline\hline
\end{tabular}
\label{tab:main-interactions}
\end{table*}

Second, Fig.~\ref{fig:main-interactions} and
Table~\ref{tab:main-interactions} give some detail about the
interactions that our scheme determines to be most relevant to fit the
TS data. The first thing to note is that we see many coincidences
among the interaction terms retained in models fitted to different TSs
(TS@10 and TS@10+300) and irrespective of the way in which the
harmonic part of $E_{\rm p}$ is treated. Another notable fact is that
the majority of terms found to be most important are {\em active} as
regards (i.e., they contribute to the energetics of) the two most
relevant structural distortions of the STO lattice, namely, the AFD
O$_{6}$-rotational instabilities of the cubic phase that lead to the
low-temperature ferroelastic structure and the low-energy polar
distortions (mostly characterized by the stretching of the Ti--O
bonds) that control the dielectric response. Finally, the
strain-phonon coupling terms always occupy preeminent positions in the
importance-ordered list provided by our automatic fit, reflecting the
known sensitivity of STO's AFD and polar phonons to strain
deformations.

\subsection{Energy boundedness}
\label{sec:boundedness2}

As mentioned in Section~\ref{sec:boundedness1}, the polynomial models
we obtain by default are likely to be unbounded from below, as there
is noting in our fitting procedure that controls the behavior of the
energy for very large distortions of the RS. In the course of this
work we have tried a number of strategies to tackle this problem, most
of which were not easy to implement in an automatic way, or led to
models with a significantly reduced accuracy. Here we briefly describe
some of those attempts, and in the end discuss the approach we found
most satisfactory and finally implemented.

\subsubsection{Quasi-automatic approaches}

Ideally we would like to have an energy-bounding strategy that is as
automatic as our model-construction procedure. Here we discuss two
representative alternatives that comply with such a requirement.

One possibility is to supplement the models with a bounding potential
that (1) is small (or even null) for the distortion amplitudes that
are typical of our TSs and (2) overpowers our fitted potential outside
that region to guarantee boundedness from below. This approach has two
distinct advantages: the resulting models are bounded by construction
and their quality, as measured by the GF used to fit them, is
essentially unaffected. However, it also presents some serious
drawbacks, at least for a material as demanding as STO: The models
thus bounded suffer from a drastic loss of predictive power for
configurations not contained in the TS (in particular, the highly
desirable matching of energy scales for MD trajectories at different
temperatures, which is shown in Fig.~\ref{fig:MD-trajectories}, is all
but lost) and tend to display spurious local minima of the energy that
compromise the results of statistical simulations. Unfortunately,
fighting against the latter problem tends to exacerbate the former,
which led us to disregard this approach.

A second possibility is to identify, and remove from the pool of
possible interactions, the terms that cause the energy
unboundedness. While this approach may seem rather drastic, it is
partly justified by the observation (see Section~\ref{sec:analysis})
that there is some degree of arbitrariness in what regards the
coupling terms selected as most relevant by our automatic procedure,
since some of them could be replaced by similar ones without any
significant loss in the model's accuracy. We proceed as follows: We
construct the best $p$-model in the usual way; we test it by running a
set of suitable Monte Carlo annealings, which allows us to identify
run-away solutions in case the model is unbounded from below; by
examining the run-away trajectory, we determine the coupling term that
dominates the energy divergence; we remove such a term from the pool
of possible interactions and determine a new best $p$-model; we repeat
the test until a bounded model is obtained. Interestingly, we find
that in some cases such a procedure was able to deliver bounded models
of excellent quality and predictive power, indicating that this may be
a viable alternative when dealing with materials that are not as
complex as STO. However, for some of the TSs and model-construction
(FM vs. EHM) strategies considered here, the procedure does not lead
to satisfactory results. Hence, we did not pursue it further in this
work.

\subsubsection{Practical approach}

Interestingly, it is actually quite easy to identify the potentially
problematic terms in our models, either automatically or by direct
inspection. With this information at hand, it is typically trivial to
identify a higher order coupling that can most effectively {\em
  control} the run-away solution, and thus include it to construct a
well-behaved model.

For example, in the case of the FM model fitted to TS@10, our analysis
shows that the optimum potential obtained by default presents run-away
solutions associated with the interactions of the Ti--O pair, mainly
driven by a third-order term of the form (Ti$_{x}$-O1$_{x}$)$^{3}$
(see term number 24 in Table~\ref{tab:full-FM-TS10}; to describe the
couplings, we adopt the compact notation used in
Appendix~\ref{sec:app2}) as well as other more complex fourth-order
terms. It is trivial to identify the sixth-order couplings [e.g. of
  the form (Ti$_{x}$-O1$_{x}$)$^{6}$] that can control the
corresponding divergences. We thus extend the optimum model by adding
these higher-order interactions, and refit the parameters to TS@10; in
this way we compute the new high-order coupling parameters and obtain
revised values for the original interactions. In the case of the FM
model fitted to TS@10, this procedure allowed us to obtain an
energy-bounded model. (Strictly speaking, as shown in
Table~\ref{tab:full-FM-TS10}, this model also required the addition of
a term of the form $\sim$(Sr$_{x}$-O1$_{x}$)$^{6}$ to control another
similar run-away solution, driven in this case by the relative Sr--O
displacements.)  Other cases were slightly more involved, but could be
resolved by e.g. supplementing the TS with a few higher-temperature
configurations (which we find tend to result in positively-defined
high-order interactions that provide boundedness from below) or
imposing by hand small and positive values for the parameters of (some
of) the high-order terms (which we find has a minor impact on the
quality of the models once the other parameters are refitted under
this constraint). Importantly, the models thus bounded continue to be
very accurate as regards both the GF value and the other set of
validation criteria discussed in Section~\ref{sec:cv-alternative}; see
Table~\ref{tab:detail} for some details on the accuracy of the optimum
energy-bounded potentials.

Finally, let us note that it might be possible to design an automatic
implementation of this kind of correction. Let us consider an
arbitrary polynomic coupling term in our potential, which we can write
as $u^{n}v^{m}...$, where $n$ and $m$ are integer numbers while $u$
and $v$ represent any possible displacement-difference or strain
factors. For any such term, it is trivial to find a related one
$u^{n'}v^{m'}...$ where the primed exponents are defined as the
smallest even number such that $n'>n$, $m'>m$, etc. If this new
coupling has a positive parameter associated to it, it will obviously
bound any run-away solution the original term may lead to. Hence, we
can automatically identify bounding interactions for each of the
couplings selected by our potential-constructions procedure; what we
still lack is an automatic way to find a positive coupling parameter
(by fitting to a high-temperature MD trajectory, or maybe
heuristically) that results in accurate and bounded models. This
development remains for future work.

\begin{figure}
\includegraphics[width=\columnwidth]{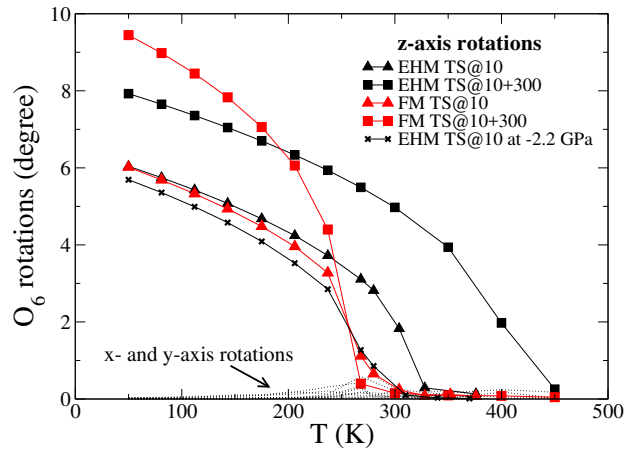}
\caption{Computed evolution of the AFD order parameter, as a function
  of temperature, for our TS@10 and TS@10+300 models. We also show the
  result for a pressure-corrected version of the EHM model fitted to
  TS@10 (see text).}
\label{fig:MC}
\end{figure}

\subsection{Simulation of SrTiO$_{3}$'s structural transition}

Let us now discuss the predictions that our different models yield for
STO's ferroelastic phase transition. To simulate STO as a function of
temperature, we run MC simulations using a periodically-repeated
12$\times$12$\times$12 supercell that contains 8640 atoms; for
temperatures close to the phase transitions, we use
16$\times$16$\times$16 supercells (20480 atoms) to better tackle
finite-size problems. Typically, the calculation at a given
temperature starts from a quasi-thermalized configuration (obtained
from an equilibrium simulation at a neighboring temperature), which we
let evolve for 20,000 MC sweeps to fully equilibrate. This is followed
by 40,000 additional sweeps to compute statistical
averages. Figure~\ref{fig:MC} shows the results thus obtained for the
TS@10 and TS@10+300 models; the figure shows the temperature
dependence of the angle characterizing the antiphase rotations of the
O$_{6}$ octahedra, which is the key order parameter to study STO's
transition.

As we can see, all the models yield a structural transformation in
which O$_{6}$ rotations about the $z$ axis occur. This is exactly the
same transition that is observed experimentally, between the
high-temperature cubic $Pm\bar{3}m$ phase and the low-temperature
tetragonal $I4/mcm$ phase; hence, all our models describe the
qualitative behavior of STO correctly. (In our simulations, the
low-symmetry phase may present O$_{6}$ rotations about any of the
principal axes in the cubic lattice; for clarity, we process our
results so that the symmetry-breaking distortion is always oriented
along $z$.)

Nevertheless, the different models lead to significant quantitative
differences as regards two important features, namely, the value of
the O$_{6}$ rotations in the limit of low temperatures and the
transition temperature $T_{\rm t}$. The former discrepancies were to
be expected, as they directly reflect the accuracy of our models to
describe the ground state structure; for example, in agreement with
the data in Table~\ref{tab:detail}, the FM model fitted to TS@10+300
renders the largest rotation angle at low temperatures; such a value
differs significantly from the smaller and nearly-identical values
that result from the models fitted to TS@10, which also agree very
well with the DFT result.

In contrast, the latter discrepancies on $T_{\rm t}$ were not obvious
to predict {\sl a priori} and provide us with new and important
information on the behavior of our models. In essence, we find a
relatively low $T_{\rm t}$ of about 270~K for the two models fitted to
TS@10, as well as for the FM model fitted to TS@10+300. In contrast,
the EHM model fitted to TS10+300 results in a transition at a much
higher $T_{\rm t} \approx 375$~K. It is instructive to compare these
findings with expectations from the literature on soft-mode driven
structural phase transitions. According to the criterion proposed in
Ref.~\onlinecite{abrahams68}, which is frequently cited and followed
by workers in the field, the transition temperature correlates
directly with the magnitude of the structural distortion in the ground
state; this rule would predict that our FM model fitted to TS10+300
should present the highest $T_{\rm t}$ among all our models, in
obvious disagreement with our findings. Nevertheless, some of us
\cite{wojdel14b} have recently revised the connection between
transition temperatures and ground state features, concluding that the
feature most directly linked to $T_{\rm t}$ is the energy difference
between the ground state and the high-symmetry structure. This new
rule would suggest that our EHM model fitted to TS10+300 should
present the highest $T_{\rm t}$ by far, while the other three models
should yield transformations at a similar temperature; this is
precisely what we find, thus supporting the conclusions of
Ref.~\onlinecite{wojdel14b}.

Let us comment briefly on the comparison of our results with
experiment. The experimental $T_{\rm t}$ of STO is known to be 110~K
\cite{unoki67}, and the O$_{6}$ rotation angle is about 2.1$^{\circ}$
at 4.2~K \cite{unoki67}. Our quantitative results severely
overestimate these two key quantities, as we obtain values of 270~K
and 6.5$^{\circ}$, respectively, for the TS@10 models. Discrepancies
of this sort, between experimental and theoretical results for STO,
are not new, and it is well accepted that they must be partly related
to LDA's overbinding error. Accordingly, previous first-principles
models of STO have been empirically corrected by introducing an
expansive external pressure \cite{wojdel13,zhong95b}; also, the DFT
results improve significantly when other functionals are used which
correct for LDA's error \cite{wahl08}. In our case, we tested this
effect by correcting our EHM model fitted to TS@10 with an hydrostatic
pressure of $-$2.2~GPa, which we numerically find is the largest value
for which this model retains STO's experimental ($I4/mcm$) ground
state (largest tensile pressures result in the stabilization of a FE
distortion, an effect that is in compatible with the phase diagram in
Ref.~\onlinecite{zhong96}). The corrected model yields the results
shown in Fig.~\ref{fig:MC}, with a new $T_{\rm t} \approx 260$~K;
additionally, for such an applied pressure we obtain a
O$_{6}$-rotation angle at 0~K of about 6.2$^{\circ}$ and an energy
difference between the cubic and ground state structures of about
10.9~meV/f.u. Hence, this correction improves the agreement with
experiment, although not very significantly.

The remaining disagreement is not so surprising when one notes that
lattice quantum effects are known to be very important in materials
like STO. Such effects generally result in a reduction of transition
temperatures and distortion amplitudes \cite{zhong95b,iniguez02},
because of the additional ways to disorder (via quantum fluctuations)
that they provide the material with. Hence, their inclusion in our
simulations should get our results closer to the experimental
values. Nevertheless, we should note that the explicit estimates made
for STO \cite{zhong96} suggest a $T_{\rm t}$ reduction of about 15~\%
when including the quantum statistics, which would result in a still
too high $T_{\rm t} \approx 200$~K in our case. Hence, these
considerations suggest that our LDA-based models may need to be
improved as regards the description of other subtle features of the
PES -- e.g., the competition between AFD and FE instabilities -- that
are known to have dramatic effects in the transition temperatures
\cite{wojdel13,zhong95b,kornev06}. A detailed analysis of such issues,
which would be far from trivial, remains for future work.

To conclude this section, let us note that, while the comparison with
experiments on STO is far from definite, our results for the phase
transition suggest several interesting conclusions. Most importantly,
they indicate that our TS@10 models may be the potentials of choice
for essentially any investigation of this compound. This is not a
trivial observation. Note that such models focus on the description of
the ground state properties and that, except for the energy difference
between ground state and RS, their fit does not include any explicit
information about the thermally-activated behavior of the
compound. Nevertheless, they render temperature-dependent results that
seem more reliable, and are closer to experiment, than those obtained
from models explicitly fitted to yield a better description of
higher-energy configurations (most notably, our EHM model fitted to
TS@10+300).

Note also that, in order to have even better models that describe well
the ground state properties (which seems mandatory) and also approach
DFT accuracy for the higher-energy configurations, we should increase
the order of our polynomial expansion. Indeed, from inspection of the
terms retained in our models (see Figs.~\ref{fig:parameters-decay} and
\ref{fig:main-interactions}, and the discussion in
Section~\ref{sec:analysis}), it is clear that our pool of parameters
contains interactions that, in terms of both spatial range and number
of bodies involved, are far more complex than what is needed to fit
the TS data. At the same time, the models have obvious difficulties to
account simultaneously for the PES associated to small (corresponding
to TS@10) and lager (captured in TS@300) RS distortions, which suggest
that the truncation at 4th order is the main limiting factor for the
quality of the models constructed in this work.

\section{Summary and conclusions}

By taking advantage of a distinctive feature of the lattice-dynamical
potentials recently introduced by some of us \cite{wojdel13} --
namely, the linear dependence of the energy on the free parameters of
the model --, we are able to design an automatic fitting scheme that
turns the calculation of the model's parameters into a simple and fast
matrix diagonalization. We thus avoid the difficulties that hamper the
construction of other lattice potentials, which are typically
non-linear in the parameters and whose fit to a training set of data
requires a numerically-costly optimization. This unique advantage,
combined with the simple Taylor-series-like form of our models --
which provides us with a well-defined strategy to improve them in a
systematic way --, allows us to sample virtually all possible
interatomic couplings during the model construction. We can thus
generate not just one model, but a family of them, of increasing
complexity and accuracy in reproducing a training set of
first-principles data. Then, a thorough cross-validation procedure,
focused on checking our models' predictive power against the
quantities that will be most critical in the simulations of specific
materials (that is, in the example we discuss, the energetics and
structure of the ground state, as well as the energetics of dynamical
simulations at various temperatures) allows us to identify models that
are simultaneously simple (thus computationally light), predictive,
and accurate.

We demonstrate our scheme with an application to ferroelastic
perovskite SrTiO$_{3}$, a material that features soft mode-driven
structural transitions and is a model compound for the investigation
of competing lattice instabilities. Hence, this is a very demanding
test case for our method, which is requested to capture subtle
dynamical features that involve a highly anharmonic potential. We show
that our scheme successfully allows us to rank the interatomic
interactions by order of importance, revealing that the main features
of the potential energy surface are controlled by couplings that are
relatively short in range. This example also allows us to discuss in
detail the main difficulty in our potential-construction scheme,
namely, that the resulting models are not guaranteed to be bounded; we
discuss several approaches to correct this problem and show that it is
possible to implement a simple and practical strategy that results in
bounded models with a negligible loss of accuracy. Finally, we show
that our models for SrTiO$_{3}$ reproduce the basic experimental facts
about the material, and briefly discuss the physics of our
automatically chosen potentials.

We thus introduce a new scheme for the construction of
lattice-dynamical models that approach first-principles accuracy and
can be constructed in an efficient quasi-automatic way. Our scheme
takes full advantage of a particular kind of potentials that have a
very simple and general form. Such potentials are restricted to treat
cases in which the lattice connectivity and topology is respected
throughout the simulation. This constant-topology condition does not
apply whenever we have formation or breaking of chemical
bonds. Nevertheless, our models are an excellent choice for the
investigation of a wealth of interesting physical phenomena -- from
the equilibrium properties of stable phases (including functional
effects such as electromechanical and dielectric responses, static or
dynamical), to non-reconstructive structural transitions (e.g., all
those driven by soft phonon modes) or thermal (transport,
electrocaloric) effects --, which underlines their interest and
potential applicability. Moreover, the models can be directly used in
schemes that include a description of the relevant electronic bands
\cite{garciafernandez16}, thus mimicking an actual first-principles
calculation. Hence, we are confident that our new
potential-construction method will be very useful in future
investigations of diverse compounds and physical phenomena.

\acknowledgements

This work is funded by the Luxembourg National Research Fund through
the Pearl (Grant No. FNR/P12/4853155) and AFR (Grant No. 9934186)
programs. We also acknowledge support from MINECO-Spain through Grant
No. MAT2013-40581-P. Some calculations were run at the CESGA
supercomputing center.

\appendix

\setcounter{table}{0}
\renewcommand{\thetable}{A\arabic{table}}

\section{Hessian analysis of the goal function}
\label{sec:app1}

Following the the notation introduced in Section~\ref{sec:formalism},
the Hessian matrix associated to the equilibrium point of the goal
function is
\begin{equation}
\begin{split}
 H_{\mu\lambda} = & \frac{2}{M_{1}} \sum_{s\tau}
 \bar{f}_{\lambda\tau}(s)\bar{f}_{\mu\tau}(s) \\
& + \frac{2}{M_{2}}
 \sum_{sa} \Omega^{2}(s) \bar{\sigma}_{\lambda a}(s)\bar{\sigma}_{\mu
   a}(s) \; ,
\end{split}
\end{equation}
where, to ease the notation, we do not indicate the dependence on the
training set. The eigenvalue problem can be written as
\begin{equation}
 H_{\lambda\mu}v_{\lambda}^i = c^iv_{\lambda}^i \; ,
\end{equation}
where $v^i$ is the $i$-th normalized eigenvectors and $c^i$ the
corresponding eigenvalue. We thus have
\begin{equation}
\begin{split}
 c^i &= \sum_{\lambda\mu}v^i_{\lambda}H_{\lambda\mu}v^i_{\mu}\\
 &=\sum_{\lambda\mu}\sum_{s} \left[ \frac{2}{M_{1}} \sum_{\tau} 
   v_{\lambda}^i\bar{f}_{\lambda\tau}(s)\bar{f}_{\mu\tau}(s)v_{\mu}^i \right.\\
&   \qquad\qquad +\left.\frac{2}{M_{2}} \Omega^{2}(s) \sum_{a}
   v_{\lambda}^i\bar{\sigma}_{\lambda a}(s)\bar{\sigma}_{\mu
     a}(s)v_{\mu}^i \right]\\
&= \sum_{s} \left[ \frac{2}{M_{1}} \sum_{\tau} 
\left( \sum_{\lambda} v^i_{\lambda}\bar{f}_{\lambda\tau}(s)
\right)^{2} \right. \\
& \qquad \left. + \frac{2}{M_{2}}\Omega^{2}(s) \sum_{a}
\left( \sum_{\lambda} v^i_{\lambda}\bar{\sigma}_{\lambda a}(s)
\right)^{2} \right] \geq 0 \; .
\end{split}
\end{equation}
Consequently, the critical manifold is necessarily a minimum (if
$c^{i}>0$ $\forall i$) or a collection of minima (if $\exists i$ such
that $c^{i} = 0$). Indeed, since we can have zero eigenvalues, the
critical manifold defined by the $\partial G/\partial \theta_{\lambda}
= 0$ condition is not necessarily of dimension zero. Instead of a
single point, we can have a minimum-$G$ line, plane, etc. in the space
of the model parameters; it immediately follows that such cases
correspond to the occurrence of linear dependences in
Eq.~(\ref{SOEcompact}).

%\vskip 2mm

\section{Full FM model fitted to TS@10}
\label{sec:app2}

We give here the complete FM model fitted to TS@10.

\begin{table}
\caption{Interactions retained in the short-range part of $E_{\rm p}$
  and $E_{\rm sp}$, corresponding to the FM model fitted to TS@10. The
  interactions are described by a polynomial coupling representing the
  whole SAT that shares the same coupling parameter. All the
  strain-phonon terms are linear in the strain and quadratic in the
  atomic displacements, and are given in eV/\AA$^{2}$. As regards the
  phonon terms, the harmonic ones are given in eV/\AA$^{2}$, 3rd-order
  ones in eV/\AA$^{3}$, and 4th-order ones in eV/\AA$^{4}$. The terms
  are given in the order in which they are selected by our automatic
  fitting procedure. Asterisks mark the terms that we introduce to
  assure the energy-boundedness of the model.}
\vskip 2mm
\begin{tabular}{c@{\hskip .5cm}lr}
\hline\hline\\[-2ex]
\# & Representative interaction & Coefficient value\\
\hline
 1 & $(\text{Ti}_y-\text{O}2_y)^2 $ & $1.548\times10^{1}\;\:\,$\\
 2 & $(\text{O}3_y-\text{Ti}_y[010]) (\text{Ti}_y-\text{O}3_y)  $ & $2.406\times10^{-1}$\\
 3 & $\eta_2 (\text{O}1_x-\text{O}2_x) (\text{O}1_y-\text{O}2_y) $ & $ 6.771$\\
 4 & $\eta_2 (\text{Sr}_x-\text{O}3_x)^2 $ & $-8.876\times10^{-1}$\\
 5 & $\eta_6 (\text{Sr}_y-\text{O}3_y[001]) (\text{Sr}_x-\text{O}3_x[001])  $ & $-2.226$\\
 6 & $(\text{Sr}_x-\text{O}1_x) (\text{Sr}_z-\text{O}3_z)  $ & $-1.314$\\
 7 & $(\text{O}3_x-\text{O}3_x[101]) (\text{O}3_x-\text{Ti}_x[110])  $ & $1.606\times10^{-2}$\\
 8 & $(\text{Ti}_x-\text{Sr}_x) (\text{Sr}_x-\text{O}2_x)  $ & $1.490\times10^{-1}$\\
 9 & $(\text{Ti}_x-\text{Ti}_x[100]) (\text{Ti}_x-\text{Sr}_x)  $ & $2.764\times10^{-2}$\\
 10 & $(\text{Sr}_x-\text{O}2_x) (\text{Sr}_y-\text{O}1_y)  $ & $4.677\times10^{-1}$\\
 11 & $(\text{Ti}_x-\text{O}2_x) (\text{Ti}_x-\text{O}1_x)  $ & $-7.076\times10^{-1}$\\
 12 & $(\text{Ti}_x-\text{Sr}_x) (\text{Sr}_y-\text{O}3_y)  $ & $1.863\times10^{-1}$\\
 13 & $(\text{Sr}_x-\text{O}2_x)^2(\text{Sr}_z-\text{O}3_z)  $ & $3.243\times10^{-1}$\\
 14 &
 \makecell[l]{$(\text{Ti}_z-\text{O}2_z)(\text{Ti}_z[0\bar{1}0]-\text{O}2_z)$\\
$\times(\text{Ti}_y[0\bar{1}0]-\text{O}2_y)$} & $-2.760$\\
 15 & $(\text{Ti}_x-\text{Ti}_x[001]) (\text{Ti}_x-\text{Ti}_x[100])  $ & $2.026\times10^{-1}$\\
 16 & $(\text{O}1_y-\text{O}2_y) (\text{O}1_y-\text{O}2_y[010])  $ & $-2.662\times10^{-1}$\\
 17 & $(\text{Ti}_y-\text{Sr}_y) (\text{Sr}_y-\text{O}2_y)  $ & $-1.878\times10^{-1}$\\
 18 & $(\text{Sr}_x-\text{O}2_x)^2(\text{O}2_x-\text{Sr}_x[001])^2 $ & $5.229\times10^{-1}$\\
 19 & $(\text{O}1_y-\text{O}3_y) (\text{Sr}_y-\text{O}1_y) (\text{O}1_x-\text{O}3_x)  $ & $-3.375\times10^{-1}$\\
 20 & $(\text{O}2_z-\text{Ti}_z[100]) (\text{O}2_z-\text{O}2_z[11\bar{1}])  $ & $-1.081\times10^{-2}$\\
 21 & $(\text{O}1_z-\text{Sr}_z[010])^2(\text{Sr}_y-\text{O}1_y)^2 $ & $9.529\times10^{-1}$\\
 22 & $\eta_1 (\text{Ti}_x-\text{O}3_x)^2$ & $-8.372$\\
 23 & $\eta_4 (\text{Sr}_z-\text{O}1_z)^2$ & $-3.203$\\
 24 & $(\text{Ti}_y-\text{O}2_y)^3 $ & $-8.814$\\
 25 & $\eta_3 (\text{O}2_x-\text{O}2_x[10\bar{1}])^2$ & $-1.791\times10^{-1}$\\
 26 & $(\text{O}1_x-\text{O}1_x[010]) (\text{O}1_x-\text{Ti}_x[010])  $ & $6.514\times10^{-2}$\\
 27 & $(\text{Sr}_y-\text{O}2_y[010]) (\text{Sr}_y-\text{Sr}_y[010])  $ & $5.390\times10^{-2}$\\
 28 & $(\text{O}1_z-\text{Ti}_z[010]) (\text{Ti}_z-\text{O}1_z)  $ & $8.515\times10^{-2}$\\
 29 & $(\text{Ti}_z-\text{O}1_z[100]) (\text{Ti}_z-\text{O}1_z)  $ & $-2.675\times10^{-1}$\\
 30 & $(\text{Sr}_z-\text{O}1_z) (\text{Sr}_y-\text{O}1_y)^3 $ & $8.967\times10^{-1}$\\
 31 & $(\text{O}2_z-\text{O}3_z) (\text{O}2_y-\text{O}3_y) (\text{Sr}_y-\text{O}2_y)  $ & $-3.477\times10^{-1}$\\
 32 & $(\text{Sr}_y-\text{O}1_y) (\text{O}1_y-\text{Ti}_y[011])  $ & $-3.635\times10^{-2}$\\
 33 & $(\text{Ti}_x-\text{O}2_x)^2(\text{O}1_y-\text{O}3_y)^2 $ & $2.332\times10^{-1}$\\
 34* & $(\text{Ti}_x-\text{O}1_x)^6 $ & $2.595\times10^{3}\;\:\,$\\
 35* & $(\text{Ti}_z-\text{Sr}_z)^6 $ & $2.434\times10^{2}\;\:\,$\\[1ex]
\hline\hline
\end{tabular}
\label{tab:full-FM-TS10}
\end{table}

Let us first introduce our notation to describe STO's RS and its
distortions; this is the notation used in this Appendix and throughout
the paper. In Cartesian coordinates, the cell is given by lattice
vectors ${\bf a} = a(1,0,0)$, ${\bf b} = a(0,1,0)$, and ${\bf c} =
a(0,0,1)$, where $a$ is the lattice constant given in Angstrom
(obtained to be 3.865~\AA\ from our LDA relaxation).  The relative
coordinates of the atoms in the cell are: $(0,0,0)$ for Sr, $(1/2,
1/2, 1/2)$ for Ti, $(0,1/2,1/2)$ for the first oxygen (O1),
$(1/2,0,1/2)$ for O2, and $(1/2,1/2,0)$ for O3. We denote the
Cartesian coordinates by $x$, $y$, and $z$; then, for example, we
write Ti$_{x}$ to denote the displacement along $x$ of the Ti atom in
the cell at the origin (i.e., this is the Ti atom located at
$a(1/2,1/2,1/2)$ in the RS); O2$_{z}$ would be the displacement along
$z$ of the O$_{2}$ atom in the cell at the origin (i.e., this is the
oxygen atom located at $a(1/2,0,1/2)$ in the RS). Finally, we use a
special notation to refer to atoms located at other lattice cells;
hence, for instance, Sr$_{y}$[110] stands for the displacement along
$y$ of the Sr atom located at position $a(1,1,0)$ in the RS. Finally,
the homogeneous deformations of the lattice are given by strains in
Voigt notation, following the standard convention.

Table~\ref{tab:full-FM-TS10} gives the full short-range part of
$E_{\rm p}$, as obtained by fitting the FM model to the TS@10 training
set described in the main text of the paper. 

To complete the description of the model, we also need to list the
values of the parameters that we compute directly from the LDA
calculation and whose corresponding couplings are included in $E_{\rm
  fixed}$ during the fits. First, we have the long-range dipole-dipole
interactions in $E_{\rm p}$, which are fully characterized by the
tensors in Table~\ref{tab:dipole}. Finally, Table~\ref{tab:elastic}
gives the LDA-computed elastic tensor.

\begin{table}[th!]
 \caption{Born dynamical effective charges in elemental charge
   units. We have a 3$\times$3 Born tensor for each atom; however,
   because of the cubic symmetry of the RS, the tensors are strictly
   diagonal and, hence, only the diagonal terms are given in the
   table. The high-frequency dielectric permittivity tensor (which
   accounts for pure electronic contributions) is isotropic and
   diagonal for STO's cubic RS, and is thus characterized by a single
   number $\epsilon_{\infty} = 6.35$.}
\vskip 2mm
 \begin{tabular}{crrr}
  \hline\hline
  & xx & yy & zz \\
  Sr & 2.55 &   2.55  &  2.55\\
  Ti & 7.33 & 7.33 & 7.33\\
  O1 & $-$5.77 & $-$2.06 & $-$2.06\\
  O2 & $-$2.06 & $-$5.77 & $-$2.06\\
  O3 & $-$2.06 & $-$2.06 & $-$5.77\\
 \hline\hline   
 \end{tabular}
\label{tab:dipole}
\end{table}

\begin{table}[th!]
\caption{Non-zero elastic constants computed for the cubic phase of
  STO, given in Voigt notation.}
\vskip 2mm
\begin{tabular}{l}
 \hline \hline
 $C_{11} = C_{22} = C_{33}$ = 388.3 GPa \\
 $C_{12} = C_{13} = C_{23} = C_{21} = C_{31} = C_{32}$ = 111.4 GPa \\
 $C_{44} = C_{55} = C_{66}$ = 118.8 GPa \\
\hline\hline
\end{tabular}
\label{tab:elastic}
\end{table}

%\bibliography{biblio}{}


\end{document}